# Understanding the influence of yttrium on the dominant twinning mode and local mechanical field evolution in extruded Mg-Y alloys

Chaitali Patil[1], Qianying Shi[1], Abhishek Kumar[2], Veera Sundararaghavan[2], John Allison[1,3]

[1]Department of Materials Science and Engineering, University of Michigan, Ann Arbor, MI 48109

[2]Department of Aerospace Engineering, University of Michigan, Ann Arbor, MI 48109

[3]Department of Mechanical Engineering, University of Michigan, Ann Arbor, MI 48109

**Abstract: T**winning constitutes one of the important deformation mechanisms for Mg alloys. This study focuses on tension twins during uniaxial compression of Mg-Y alloys, with three key aspects: the orientation specificity of twin grains; the relative evolution of critical resolved shear stresses with increasing Y content; and the local stress and strain evolution at twin locations. We conducted both experimental characterization and crystal plasticity analysis for this purpose. In Mg-7 wt%Y alloy, we observed the presence of TT2-$\{11\bar{2}1\}$ tension twins along with the commonly observed TT1-$\{10\bar{1}2\}$ twins. A higher Y concentration in Mg decreased the formation of TT1 twins and promoted the formation of TT2 twins. A previously unreported group of crystallographic orientations with a higher global Schmid factor for <c+a> slip is identified, which exhibits TT1 twinning with increasing compression strain. To elucidate the effect of Y content on twin activity and the evolution of local mechanical fields, both TT1 and TT2 tension twin modes were incorporated into PRISMS-Plasticity, an open-source, finite element-based crystal plasticity solver. Simulations of the deformation of four binary Mg-Y alloys under compression were performed. A comprehensive statistical analysis was conducted to examine the correlations among initial orientations, stress-strain distributions, and the activities of both tension twin types as functions of Y concentration. The plasticity analysis revealed that the ratio of the critical resolved shear stress for prismatic and pyramidal slip to the TT1 twin decreased with Y addition. Conversely, the slip-to-twin CRSS ratio for TT2 increased, thereby serving as a potential indicator of differential twin activity with Y addition in Mg alloys. Additionally, even at small volume fractions, higher strain accumulation was predicted locally at TT2 twin sites in PRISMS-Plasticity simulations. The mean equivalent strain at TT2 twin locations was higher than the average values for the representative volume element (RVE) and TT1 twins, suggesting potential influence on localized phenomena such as recrystallization or twin nucleation. These insights will help in understanding the local mechanical properties of Mg alloys, thereby facilitating their design and utilization in advanced engineering applications.

## 1. Introduction

Magnesium alloys, with properties such as low density, high specific strength, and high elastic modulus, can provide economical and environmentally friendly alternatives in transportation, aerospace, and energy applications. [1,2]. To further improve the strength, Mg alloys with rare-earth elements (RE) additions have been developed [3,4]. RE alloying additions, like yttrium (Y), weaken the strong basal texture formation during manufacturing and improve the formability of Mg [5,6]. It was also reported that the Y addition benefits the alloy's ductility [7,8], creep resistance [9,10] and corrosion resistance [11]. While the most notable effect of Y addition in Mg is activating non-basal slip, such as pyramidal $<c+a>$ dislocations [12–14], the influence of Y on twinning behavior is also of great interest.

Twin formation significantly affects the mechanical behavior of Mg by accommodating deformation along the c-axis. **Figure 1** presents a schematic of two types of tension twins, TT1 twins – $\{10\bar{1}2\} <10\bar{1}\bar{1}>$ and TT2 – twins $\{11\bar{2}1\} \langle\bar{1}\bar{1}26\rangle$. Both twin types accommodate tensile strain along the $<c>$ axis during the deformation. In HCP metals, the TT2 twin is the only twinning mode where all lattice points can be displaced to the correct twin position without requiring additional shuffle of atoms along the shear direction [15]. For the TT1 twins, the misorientation angle distribution shows a characteristic peak at 84.8° (±5°) with a rotation axis of $<11\bar{2}0>$ [16]. For the TT2 -$\{11\bar{2}1\}$ twins, the misorientation angle distribution shows a characteristic peak at 35.1° (± 5°) with a rotation axis of $<10\bar{1}0>$ [16]. The twin shear for the TT1 is given by $\frac{(R^2-3)}{R\sqrt{3}}$, where $R = \frac{c}{a}$, while the twin shear for the TT2 is given by $\frac{1}{R}$ [17]. Thus, for Mg alloys, the characteristic twin shear for the TT1 is ≈0.129 while the characteristic twin shear for TT2 is ≈0.616 considering a c/a ratio for Mg of 1.624. Hence, the shear accommodation by TT2 twins is nearly five times that of TT1, which can influence both strain accommodation and twin evolution during deformation.

TT1 twins are commonly observed in Mg alloys. Extensive characterization has been conducted to understand the nucleation, propagation, and twin interactions of the TT1 twins during deformation. Although the TT2 twin is observed in HCP metals like Co [18], Ti alloys [19], and Zr alloys [20,21], it is less commonly observed in the Mg alloys. Nonetheless, observations of the TT2 twins have been reported in various Mg-RE alloys such as WE54 [22], WE43 [23], Mg-Gd-Y-Zr [24], and Mg-17wt.%Gd [25]. Specifically with Y addition, the TT2 type twinning was observed in Mg-9wt.%Y [26] and Mg-10wt.%Y [27].

With the emerging research interest in TT2 twin formation, many state-of-the-art experimental techniques, as well as modeling methods, have been employed to understand TT2 twinning in the Mg alloys. For example, using TEM/STEM analysis, Chen et al studied the twin-twin interactions of the TT2 twin variants in WE43 [23], while Zhang et al studied the TT2 activity at a higher strain rate in Mg-Gd-Y-Zr alloy [24]. Li et al. observed TT2 twin activation prior to the activation of TT1 twins in solutionized Mg-Gd-Y alloy; however, the sequence of twin occurrence was reversed in alloys subjected to additional aging treatment [28]. Gengor et al. calculated the Generalized Planar Fault Energy (GPFE) for TT2 twin variants in HCP materials using ab-initio simulations and emphasized the nonzero shuffle distortions normal to the shear plane for the TT2 twin [29]. The authors reported a nonlinear positive correlation between the elastic stiffness coefficient $C_{44}$ and twin boundary energy. Thus, most of the recent research has been focused on understanding twin energetics or twin activation in different heat-treated or deformation conditions. However, a detailed analysis of the TT2 twinning activity and its influence during deformation in Mg alloys is still lacking in the literature. Hence, the current study employs detailed experimental characterization and crystal plasticity analysis to gain further insights into correlations between the TT2 twin activity and micromechanical field evolutions during the deformation.

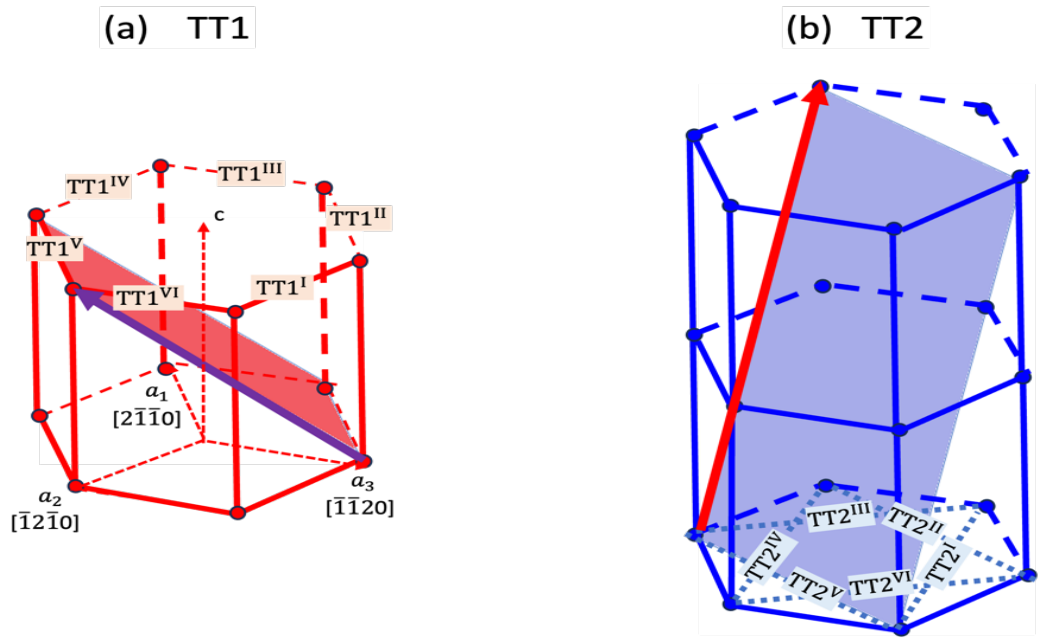

**Figure 1.** Schematic representation of (a) the $\{10\bar{1}2\}$ (TT1) twin and (b) the $\{11\bar{2}1\}$ (TT2) twin geometry in the HCP crystal system. One of the invariant shear planes, $\mathbf{k_1}$, and the shear direction $\boldsymbol{\eta_1}$ for both twin types are highlighted in the schematic. Projections of the twin plane of remaining variants on the basal plane are also denoted.

Very few studies in the literature have employed crystal plasticity for Mg-Y alloys. For example, Li et al. conducted crystal plasticity finite element (CPFE) analysis for Mg-0.8wt.%Y, which showed higher non-basal slip activity and lower twin activity at higher strains [30]. Huang et al utilized CPFE analysis to understand the anomalous TT1 twin activity in Mg-2wt.%Y alloy [31], while Su et al. utilized nanoindentation and CPFE of Mg-2wt.%Y alloys for analyzing orientation dependence during the deformation [32]. Kula et al. illustrated a monotonic increase in the critical resolved shear stresses (CRSS) for active deformation modes with increasing Y addition [33]. These investigations all involved alloys with relatively low Y concentrations; thus, the TT2 twinning mode was not considered in the CPFE simulations. Recently, Siska et al provided insights into stress distribution in the single crystal with an ellipsoid twin geometry [34]. However, polycrystalline deformation can have additional influences due to the orientations of the constituent grains and their grain neighborhoods. Therefore, the current work presents a systematic statistical analysis of polycrystalline deformation with both TT1 and TT2 twin activation in Mg-Y alloys, using crystal plasticity and experimental twin characterization. Such experimental characterization, complemented by CPFE analysis of twinning, can provide a broader spectrum of insights into the evolution of the local mechanical field and twinning activity during polycrystalline deformation.

In this work, we employed both experimental characterization and crystal plasticity analysis of Mg-Y alloys to address three key questions: 1. Which orientations favor twin activation in Mg-Y alloys, and does the orientation preference change with increasing strain level? To explore this, the orientation specificity of both twin types is experimentally analyzed. 2. How does increasing Y content affect the CPFE-assessed critical resolved shear stresses of different slip and twin modes, providing insights into twin propensity in Mg alloys with varying Y levels? For this purpose, the open-source crystal plasticity framework PRISMS-Plasticity [35–37] is utilized to simulate the deformation of Mg alloys with four different Y contents. A representative volume element (RVE), based on the experimental extrusion texture, is employed for all simulations. Assessed CRSS values and other constitutive parameters are calibrated against experimental flow data and twin evolution. 3. Lastly, how does different twin activity influence local stress and strain distributions during deformation of Mg-Y alloys? Since all simulations use the same microstructure, observed differences in the local field evolutions can be attributed to variations in deformation activity, instead of differences in the initial orientations. Given the growing importance of Mg-RE alloys, this work provides valuable insights into how solute yttrium influences active twinning modes and the local mechanical fields during compression of extruded Mg-Y alloys.

## 2. Methods

### 2.1. Materials and experimental procedure

Two Mg-Y alloys containing different levels of Y concentrations, 1 wt.% (0.28 at.%) and 7 wt.% (2.02 at.%), are used in this study. These alloys are referred to as Mg-1Y and Mg-7Y, respectively, throughout the remainder of the text. These materials were provided by CanmetMATERIALS, Canada in the form of extruded bars. The alloy composition and extrusion procedure were detailed in a previous publication [38]. The as-extruded bars with a diameter of 15 mm were sectioned and then annealed in a Thermolyne box furnace at 400 °C for 24 hours, followed by water quenching.

Cylindrical specimens with a geometry of 10 mm in diameter and 12 mm in height were machined from materials after the above annealing heat treatment using electrical discharge machining (EDM). The longitudinal direction of cylindrical specimens was aligned with the extrusion direction, which is also parallel to the compression axis. The compression experiment was performed using an Instron load frame equipped with a 100 kN load cell at a constant displacement rate of 0.03 mm/min. Three levels of total strain were applied, 3%, 5%, and 8% for each alloy. It should be noted that two ends of the cylindrical specimens were hand ground to an 800 SiC grit-finish, and Teflon tape was applied between the two surfaces to reduce the friction between specimens and compression dies. Additionally, the strain measurement represented the macroscopic average strain of the testing specimen.

After compression, the cylindrical specimens were sectioned parallel to the loading axis for microstructure characterization. Metallurgical samples were prepared using standard grinding and polishing practices for Mg alloys until the final polish step with 1 μm diamond paste. Buehler MetaDi fluid was used as a polishing lubricant instead of water. An acetic-nitric solution (60 mL ethanol, 20 mL water, 15 mL glacial acetic acid, and 5 mL of nitric acid) was used to etch specimens for approximately 15 seconds.

To reveal the crystallographic orientation of grains and twins, electron backscatter diffraction (EBSD) scanning was conducted using a Tescan Mira3 electron microscope equipped with an EDAX Velocity camera. Each EBSD scan was performed on a 667 μm long and 500 μm high region with a step size of 500 nm at a voltage of 30 kV and a working distance of approximately 20 mm. Four EBSD scans were obtained from the sample center regions for each compression condition for a consistent and statistical comparison. The collected EBSD raw data were analyzed using either OIM software or MTEX, which is a free and open-source MATLAB toolbox for analyzing and modeling crystallographic orientations [40].

The initial grain structure and texture for alloys Mg-1Y and Mg-7Y are shown in **Figure 2** (a) and (b), respectively. Grains in both alloys are fully recrystallized after the annealing heat treatment at 400 °C for 24 hours. A slight difference in grain size was observed between these two materials: the grain size in the Mg-1Y was 51±18 μm, whereas it was 32±11 μm in alloy Mg-7Y. A weak basal fiber texture was observed in both annealed alloys, which was inherited from the extruded bar microstructure [38] . As indicated by the texture analysis shown in the pole figures in **Figure 2** (a) and (b), the <0001>-*c* axis tended to align with the radial direction and perpendicular to the extrusion direction.

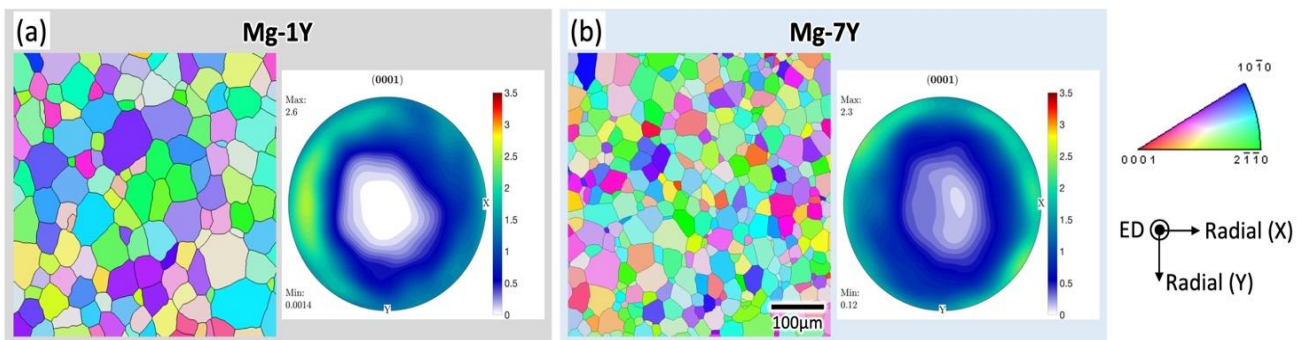

**Figure 2.** Microstructure shown in EBSD-IPF maps and texture analysis shown in pole figures for (a) Mg-1Y and (b) Mg-7Y alloys in the annealed condition. The color in IPF maps represents the mean orientation of each grain viewed from the extrusion direction.

**2.2. Crystal plasticity simulation methodology**

Crystal plasticity finite element (CPFE) simulations were performed using the open-source PRISMS-Plasticity framework [36,41]. The utility of PRISMS-Plasticity has been demonstrated in various applications, such as fatigue simulations [42,43], texture evolution [44,45], and the effects of grain size on flow behavior [46]. Details of the finite strain framework and its implementation can be found in [36]. In this work, the rate-independent constitutive formulation [36] is employed for the deformation simulations. Here, we will summarize the constitutive formulation.

2.2.1 **Crystal plasticity model**

Assuming the multiplicative decomposition, the deformation gradient **F** can be written in terms of the elastic ($\mathbf{F}^e$) and plastic deformation gradient ($\mathbf{F}^p$) as,

$$\mathbf{F} = \mathbf{F}^e \mathbf{F}^p \tag{1}$$

The velocity gradient in the deformed configuration (**L**) is given by $\dot{\mathbf{F}} \mathbf{F}^{-1}$. Thus, using the above equation, the velocity gradient, **L**, can be written as,

$$\mathbf{L} = \dot{\mathbf{F}}^e \mathbf{F}^{e-1} + \mathbf{F}^e \dot{\mathbf{F}}^p \mathbf{F}^{p-1} \mathbf{F}^{e-1} \tag{2}$$

Accordingly, the total velocity gradient tensor can be additively decomposed into the elastic and plastic components, i.e.,

$$\mathbf{L} = \mathbf{A} + \boldsymbol{l}^p \tag{3}$$

where $l^e = \dot{\mathbf{F}}^e\mathbf{F}^{e-1}$ and $l^p = \mathbf{F}^e\dot{\mathbf{F}}^p\mathbf{F}^{p-1}\mathbf{F}^{e-1}$ are in the deformed configuration. The $\dot{\mathbf{F}}^p\mathbf{F}^{p-1}$ in the definition of $l^p$ represents the plastic velocity gradient in the intermediate configuration ($\mathbf{L}^p$). The $\mathbf{L}^p$ can also be determined in terms of the superposition of the shear rates on multiple slip systems. In this work, twinning is considered as a pseudo-slip system, and hence, $\mathbf{L}^p$ is determined in terms of total shear due to slip and twinning during plastic deformation. Thus, $\mathbf{L}^p$ is defined as,

$$\mathbf{L}^p = \dot{\mathbf{F}}^p\mathbf{F}^{p-1} = \sum_{\alpha=1}^{N_s+N_t} \dot{\gamma}^\alpha \mathbf{S}^\alpha \qquad (4)$$

$$\mathbf{S}^\alpha = \mathbf{m}^\alpha \otimes \mathbf{n}^\alpha \qquad (5)$$

Here, $\dot{\gamma}^\alpha$ is the shear rate on the slip system $\alpha$ and $\mathbf{S}^\alpha$ is the Schmid tensor in the intermediate configuration. $N_s$ is the number of slip systems, and $N_t$ is the number of twin systems. $\mathbf{m}^\alpha$ and $\mathbf{n}^\alpha$, respectively, are the unit vectors in the slip direction and slip plane normal for slip systems in the intermediate configuration. Similarly, in the case of twin systems, $\mathbf{m}^\alpha$ and $\mathbf{n}^\alpha$, respectively, represent the twin shear direction and twin plane normal. In the rate-independent formulation, yield will occur if the resolved shear stress on the slip system $\alpha$, $\tau^\alpha$, reaches a critical resolved shear stress $s^\alpha$. i.e. yield surface $g^\alpha$ can be expressed as,

$$\mathbf{g}^\alpha = \tau^\alpha - s^\alpha \qquad (6)$$

The slip system is inactive if the $\tau^\alpha < s^\alpha$ or $|\tau^\alpha| = s^\alpha$, while shear accumulation is allowed if $\tau^\alpha > s^\alpha$. The resolved shear stress on the slip system is given by,

$$\tau^\alpha = \boldsymbol{\sigma} : \mathbf{S}^\alpha \qquad (7)$$

where $\boldsymbol{\sigma}$ is the Cauchy stress tensor, and : The operator denotes the standard inner product of tensors. The $s^\alpha$ evolves with the following isotropic hardening model,

$$\dot{s}^\alpha = \sum_\beta h^{\alpha\beta} |\dot{\gamma}^\beta| \qquad (8)$$

where $h^{\alpha\beta}$ are the hardening moduli. The hardening matrix h defines the influence of the slip activity on the slip system $\beta$ on the slip resistance of the slip system $\alpha$. The hardening moduli, $h^{\alpha\beta}$, are obtained using a power-law relationship as below:

$$h^{\alpha\beta} = h_0^\beta \left[1 - \frac{s^\beta}{s_s^\beta}\right]^{a^\beta} \quad if\ \alpha = \beta\ (coplanar\ systems)$$

$$= h_0^\beta q \left[1 - \frac{s^\beta}{s_s^\beta}\right]^{a^\beta} \quad if\ \alpha \neq \beta\ (non-coplanar\ systems) \tag{9}$$

where $q$ is the latent hardening ratio, $h_0^\beta$ denotes the hardening parameter for the slip system $\beta$, $s_s^\beta$ is the slip resistance at hardening saturation for the slip system $\beta$, and $a^\beta$ is a material constant for the slip system $\beta$ which governs the sensitivity of the hardening moduli to the slip resistance.

Finally, twin volume fraction evolution on twin system $k$ is defined as,

$$\dot{f}^k = \frac{\dot{\gamma}^k}{S^k} \tag{10}$$

Where, $S^k$ represents characteristic twin shear for twin system $k$.

### 2.2.2 Deformation modes and elastic anisotropy

During uniaxial deformation of Mg-Y alloys, basal <a>{0001}⟨11$\bar{2}$0⟩, prismatic ⟨a⟩{10$\bar{1}$0}⟨1120⟩, pyramidal <a>{10$\bar{1}$1}⟨11$\bar{2}$0⟩ and pyramidal ⟨c+a⟩{11$\bar{2}$2}⟨11$\bar{2}$3⟩ are activated [47,48]. Hence, all the above slip systems are considered during the deformation. Another important modification is the consideration of the TT2-{11$\bar{2}$1} twin along with the TT1-{10$\bar{1}$2} twin. In addition to the twin plane and shear direction, the characteristic twin shear for these two types of twins is also different. Given the differences in the characteristic shear values, the same amount of shear strain can be accommodated by a much smaller volume fraction of the TT2 twin than by the TT1 twin. Although the twin shear strain depends on the c/a ratio, synchrotron radiation measurements in Mg-3Wt%Y have shown only a small decrease (~0.0008) in the c/a ratio [12] and thereforesubstantial variations in twin shear due to Y addition are unlikely. Consequently, the same twin shear was considered for all alloys.

The elastic coefficients for the Mg-Y alloys were calculated based on linear regression coefficients ($\delta$) determined by Ganeshan et al [49] using first-principles calculations as,

$$\delta = \frac{C_{Mg} - C_{Mg-x}}{\Delta x} \qquad (11)$$

Here, $C_{Mg}$ and $C_{Mg-x}$ represent the elastic coefficients of the Mg and Mg-X alloys, respectively, while $\Delta x$ denotes the composition in at.%. Using Eq. 11, calculated elastic constants for the Mg-Y alloys considered in the plasticity simulations are given in Table 1. It can be observed from the anisotropy ratios of the Mg-Y alloys that the elastic anisotropy of the alloys increases with increasing Y addition from Mg-1Y.

**Table 1.** Elastic coefficients for Mg-Y alloy based on the first-principles calculations by Ganeshan et al [49].

| Alloy (Mg-Wt% Y) | C11 | C12 | C13 | C33 | C44 | Anisotropy factor |
|---|---|---|---|---|---|---|
| Mg | 59.30 | 25.70 | 21.40 | 61.50 | 16.40 | 0.98 |
| Mg-1Y | 59.70 | 25.43 | 21.23 | 61.65 | 16.03 | 0.94 |
| Mg-5Y | 61.34 | 24.45 | 20.56 | 62.26 | 14.51 | 1.27 |
| Mg-7Y | 63.36 | 23.22 | 19.73 | 63.02 | 12.64 | 1.59 |
| Mg-10Y | 63.55 | 23.10 | 19.66 | 63.09 | 12.47 | 1.62 |

2.2.3 **Initial texture of simulation RVE and CRSS calibration**

The stress-strain response of Mg-Y alloys is calibrated using the representative volume element (RVE) with 4096 grains. The (0001), ($\bar{1}010$) and ($\bar{2}110$) pole figures for the RVE are shown in **Figure 3**. As shown in the figure, the RVE texture closely matches the initial experimental texture, shown in **Figure 2**. Each grain in the RVE is modeled by a single eight-node linear hexahedral element in a 16×16×16 finite element cubic mesh.

Critical resolved shear stresses and hardening parameters were calibrated using the initial RVE under a symmetric boundary condition, with iterative trials to optimize calibration across all alloys. Average flow stress evolution and the twin fraction evolution were utilized for this purpose. As the range of alloys is calibrated, a perfect fit for each alloy can lead to overfitting for some alloys and underfitting for others. Therefore, it is ensured that the CRSS values are monotonically increasing with the Y concentration and that the stress and twin evolution are well captured by the simulated RVE for all of the alloys in this investigation. We note that experimental boundary conditions can increase heterogeneity in deformation behavior, leading to greater strain localization. However, CRSS calibration is considered an average mesoscale representation of the lattice's resistance to deformation and twin evolution. Although CRSS encapsulates the resulting material-specific flow behavior, CRSS changes cannot distinguish the governing microscopic mechanisms. Local mechanical boundary deviations from applied global strain are not accounted for in the current study.

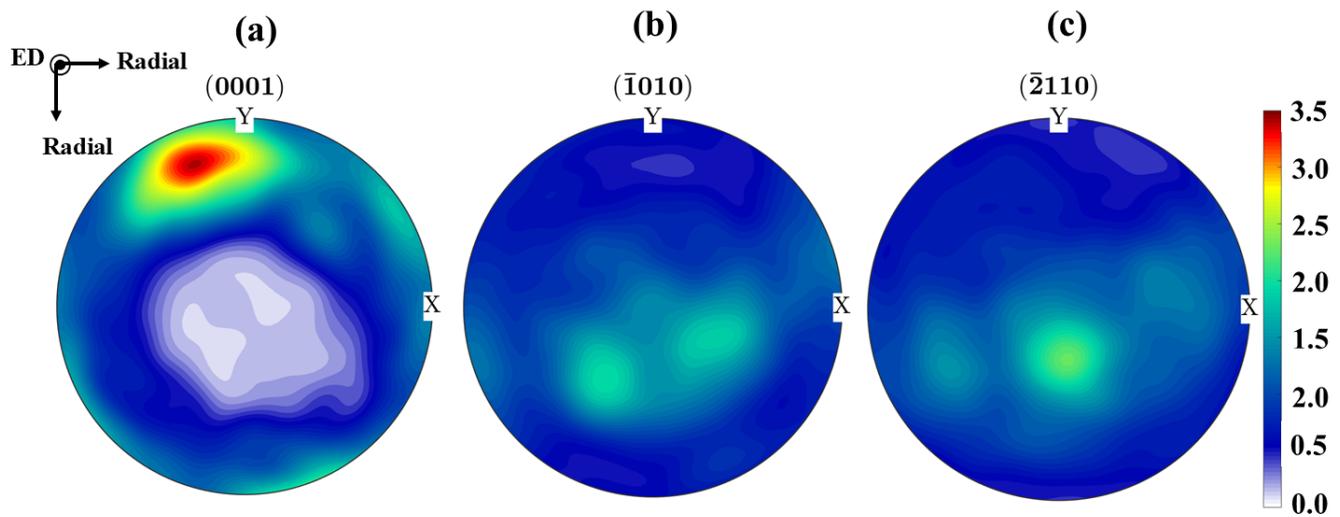

**Figure 3.** Pole figure for RVE used for the deformation simulation of Mg-Y alloys based on the initial experimentally determined texture.

## 3. Results and Discussion

### 3.1. Twin formation during compression

**Figure 4** shows the representative EBSD-IPF maps from the post-mortem EBSD characterization for each compressed condition of the two alloys. For the same strain levels, significantly less twin formation is observed in alloy Mg-7Y compared to Mg-1Y, as shown in **Figure 4**.

Although TT1 twins were observed in both alloys, increasing Y content was observed to reduce the formation of TT1 twins. **Figure 5** compares the area fraction of TT1 twins as well as the number fraction of TT1-twinned grains in two alloys as a function of the compression strain level. **Figure 5** (a) shows a significantly lower area fraction of the TT1-twinned grains in Mg-7Y samples compared to Mg-1Y samples for all tested strain levels. We also note that the rate of increase in area fraction in TT1 twinning with strain is higher in Mg-1Y compared to Mg-7Y alloys. Meanwhile, the number fraction of grains that had TT1 twins in Mg-7Y was also substantially lower than that in Mg-1Y (**Figure 5** (b)). This indicates that higher Y addition appears to retard the nucleation and/or growth of the TT1 twins.

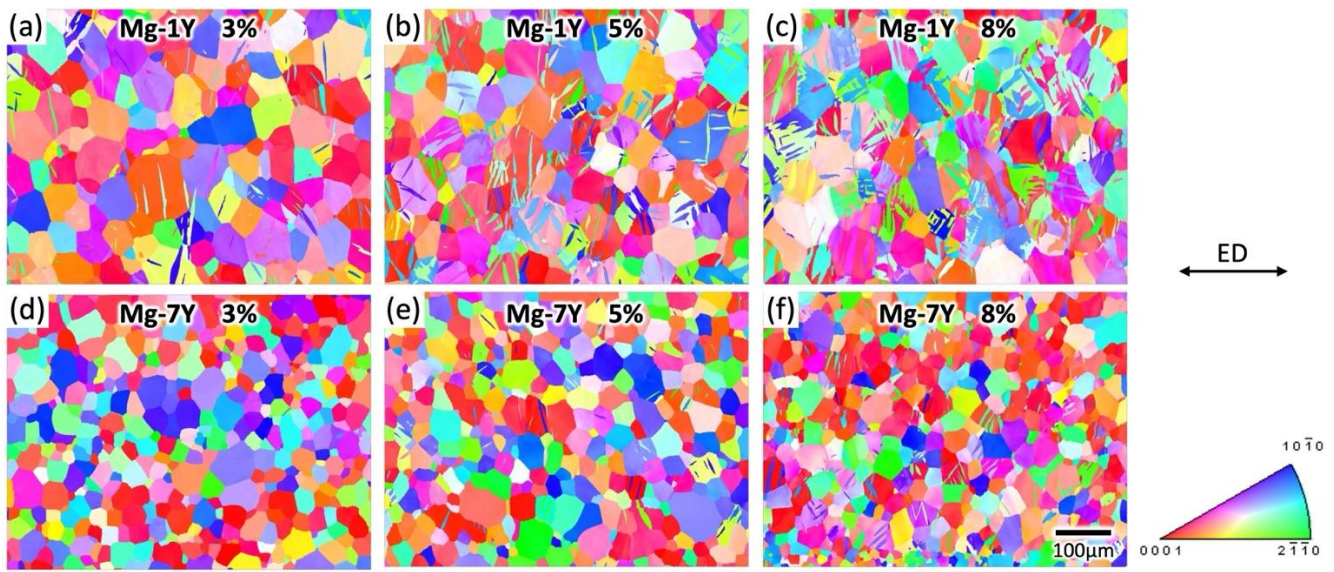

**Figure 4.** EBSD-IPF maps showing the microstructure of two alloys after compression with different strain levels. (a) Mg-1Y with 3% total strain; (b) Mg-1Y with 5% total strain; (c) Mg-1Y with 8% total strain; (d) Mg-7Y with 3% total strain; (e) Mg-7Y with 5% total strain; (f) Mg-7Y with 8% total strain. All images share the same scale bar. The compression loading direction was parallel to the extrusion direction, which is horizontal in the IPF maps shown here.

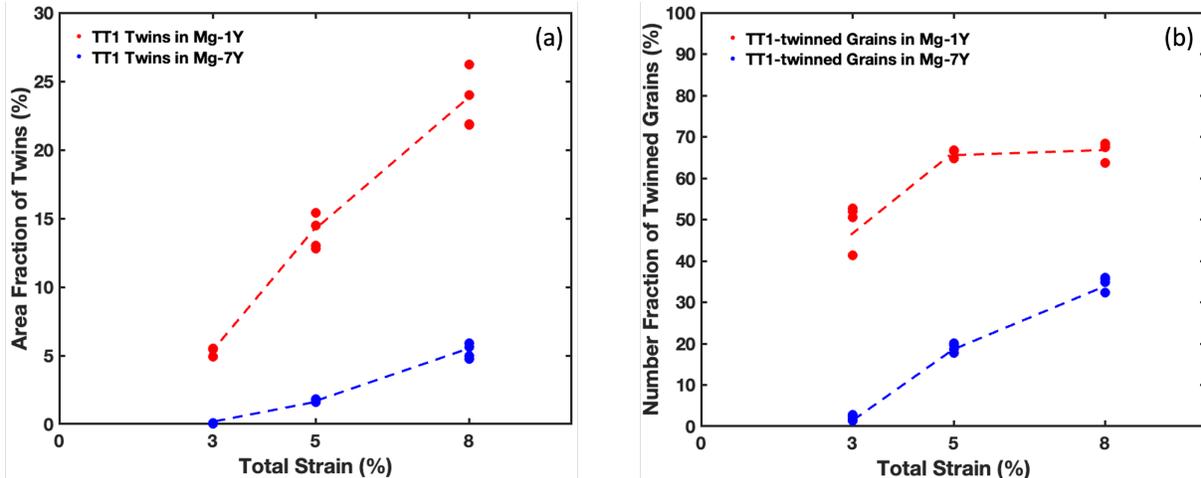

**Figure 5.** Comparison of the TT1 twin formation in two alloys with different Y concentrations. (a) Area fraction of the TT1 twins as a function of the compression strain level; (b) Number fraction of the TT1-twinned grains as a function of the compression strain level. Trendlines were added for visual assistance.

TT2 twins were observed in alloy Mg-7Y, but were rarely found in alloy Mg-1Y, suggesting that a higher Y concentration in Mg promotes the formation of TT2 twins. The amount and morphology of the TT2 twins in alloy Mg-7Y were characterized systematically and statistically. The boundaries between parent grains and TT2 twins are outlined in white lines in **Figure 6** (a), which is an EBSD-IPF map acquired for alloy Mg-7Y after 8% compression strain. It is interesting to note that TT2 twins appeared longer and thinner than TT1 twins, as indicated by the appearance of outlined TT2 twins with sharp and straight boundaries shown in **Figure 6** (a) and the aspect ratio comparison shown in **Figure 6** (b). The area fraction of TT2 twins and the number fraction of TT2-twinned grains in Mg-7Y samples are presented in **Figure 6** (c). A significantly lower fraction of TT2 twins formed in Mg-7Y compared to the TT1 twins (**Figure 5**). The area fraction of TT2 twins was 30~50 times lower than that of TT1 twins, and the number fraction of TT2-twinned grains was 20~30 times lower than that of TT1-twinned grains. As shown in Figure 6c, the amount of TT2 twins was strain-dependent and increased with increasing applied strain. At the lowest strain of 3% TT2 twins were not observed.

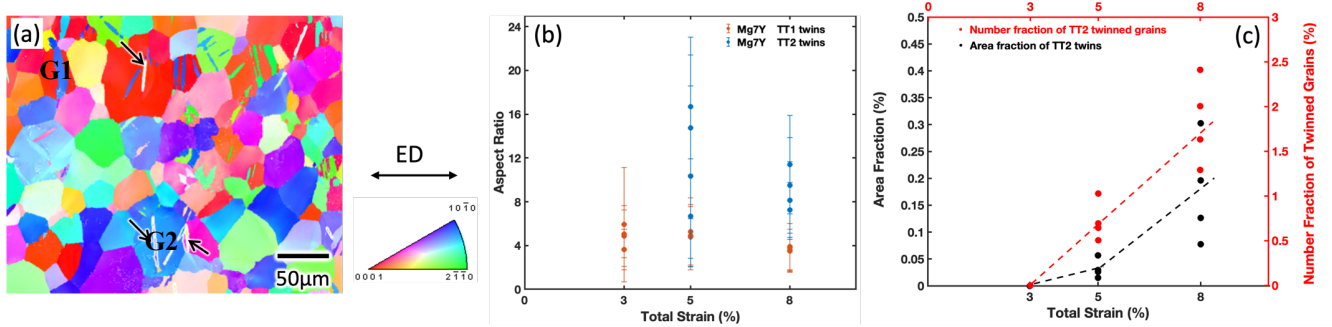

**Figure 6.** (a) EBSD IPF map showing the formation of TT2 twins in Mg-7Y after being compressed to a total strain of 8%; (b) Aspect ratio comparison between the two types of twins in Mg-7Y samples; (c) Area fraction of TT2 twins and number fraction of TT2-twinned grains in Mg-7Y samples. Trendlines were added for visual assistance. Grains G1 and G2 denoted in (a) are used for CPFE analysis in Sec.3.3.3

## 3.2 Crystallographic orientation analysis on twinned grains

### 3.2.1 TT1-twinned grains

The crystallographic orientations of all TT1-twinned grains from three strain levels of compression for both alloys were analyzed and compared with the initial microstructures. **Figure 7** (a) and (d) are the processed orientation distribution function (ODF) shown in the form of inverse pole figures. Serving as the referenced grain orientation distribution, both figures represent the overall alloy texture in the initial condition for alloy Mg-1Y and Mg-7Y, respectively. **Figure 7** (b) and (e) present the orientation distribution of all TT1-twinned grains for Mg-1Y and Mg-7Y, respectively. Accordingly, **Figure 7** (c) and (f) are the corresponding processed ODF analysis results. These results suggested there were significant changes in the orientation distribution between TT1-twinned grains and initial nondeformed grains. This can be easily noticed by comparing **Figure 7** (a) and (c) or **Figure 7** (d) and (f), where the ODF intensity peak shifted from the corner of the crystallographic orientation ⟨2$\bar{1}\bar{1}$0⟩ to the corner of ⟨10$\bar{1}$0⟩. The TT1-twinned grains predominantly had orientations close to ⟨10$\bar{1}$0⟩, as shown in **Figure 7** (b) and (e). This group of TT1-twinned grains was categorized as grain group A. In addition, it can be observed that many TT1-twinned grains were distributed around the corner of <0001>. This second group of TT1-twinned grains was categorized as grain group B.

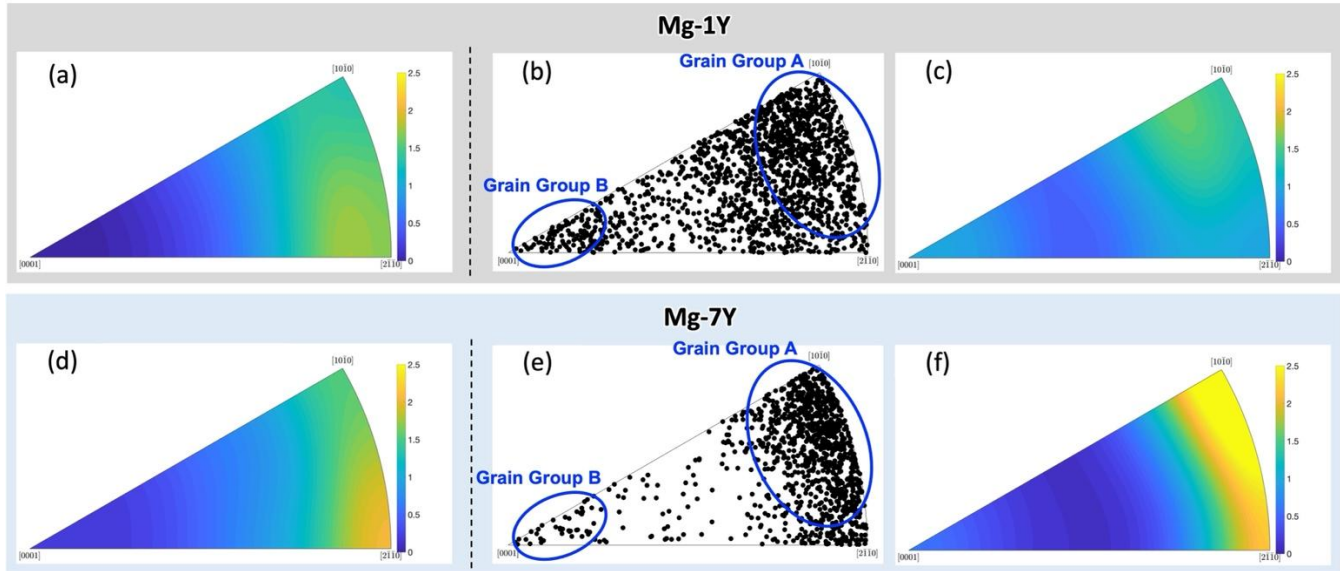

**Figure 7.** Comparison of the grain orientation distribution between grains in initial microstructures and TT1-twinned grains, which were analyzed on compressed samples with all different strain levels, for two alloys. (a) Processed ODF analysis of grain orientation distribution (alloy texture) of Mg-1Y in the initial condition, as shown in the form of inverse pole figure; (b) Orientation distribution of all TT1-twinned grains characterized for Mg-1Y; (c) Processed ODF analysis of orientation distribution in (b), as shown in the form of inverse pole figure; (d) Processed ODF analysis of grain orientation distribution (alloy texture) of Mg-7Y in initial condition, as shown in the form of inverse pole figure; (e) Orientation distribution of all TT1-twinned grains characterized in Mg-7Y; (f) Processed ODF analysis of orientation distribution in (e), as shown in the form of inverse pole figure. All inverse pole figures are generated for the compression direction (i.e. extrusion direction).

The orientation distribution of TT1-twinned grains was found to change with the level of compression strain. **Figure 8** shows the orientation distribution of TT1-twinned grains for both alloys after compression to different strain levels. With increasing compression strain, more grains belonging to group B exhibited TT1 twin formation, as observed either from **Figure 8** (a) to (c) for Mg-1Y or from **Figure 8** (d) to (f) for Mg-7Y. In addition, it should be noted that the crystallographic orientations for TT1-twinned grains in group A and group B could result in activation of different twinning systems and slip systems. Therefore, global (nominal) Schmid factor (SF) calculations were performed for different deformation systems and overlaid with the TT1-twinned grain orientations.

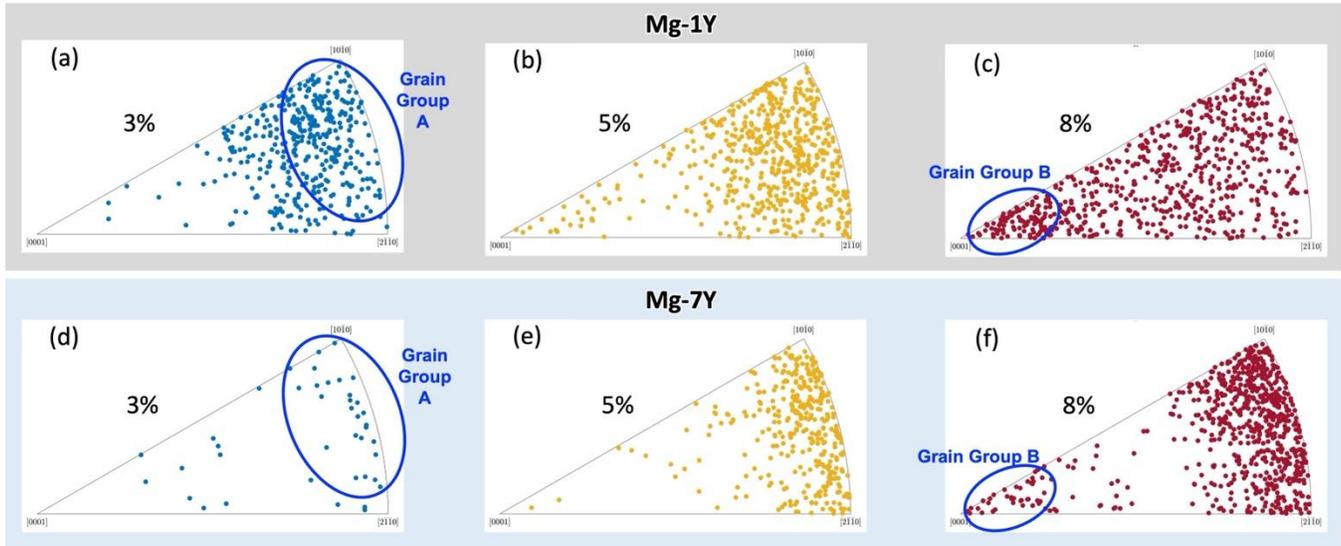

**Figure 8.** Orientation distribution of TT1-twinned grains in Mg-1Y and Mg-7Y after compression with different strain levels. (a) TT1-twinned grains in the 3% strain compressed Mg-1Y sample; (b) TT1-twinned grains in the 5% strain compressed Mg-1Y sample; (c) TT1-twinned grains in the 8% strain compressed Mg-1Y sample; (d) TT1-twinned grains in the 3% strain compressed Mg-7Y sample; (e) TT1-twinned grains in the 5% strain compressed Mg-7Y sample; (f) TT1-twinned grains in the 8% strain compressed Mg-7Y sample. All inverse pole figures are generated for the compression direction (*i.e.* extrusion direction).

**Figure 9** presents these results for compressed Mg-7Y samples at all strain levels with SF calculation for the TT1 twinning system $\{10\bar{1}2\}\langle\bar{1}011\rangle$, pyramidal I <c+a> slip system $\{10\bar{1}1\}\langle11\bar{2}3\rangle$, and pyramidal II <c+a> slip system $\{11\bar{2}2\}\langle11\bar{2}3\rangle$. Grains from group A had higher SF for the TT1 twinning system, and grains from group B had higher SF for the pyramidal slip system. It suggests that grains from group A are potentially easily twinned due to their favorable orientations for the TT1 twinning system. With increased compression strain levels, more grains from group B activated TT1 twins, suggesting the possible association between pyramidal slip and TT1 twin nucleation due to the particular orientation of grains in group B.

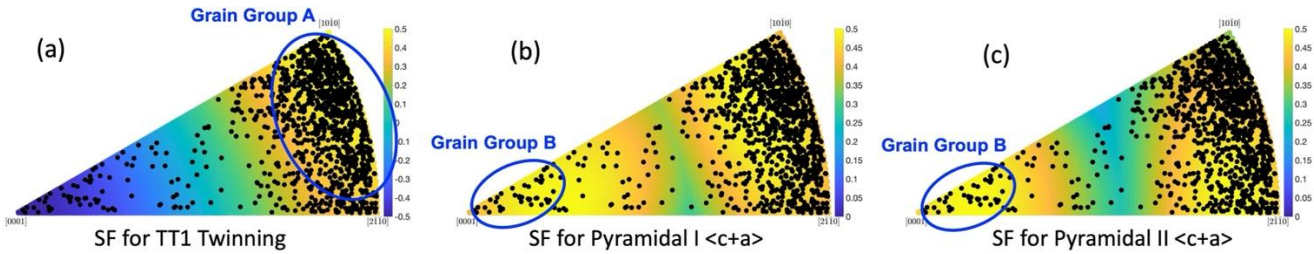

**Figure 9.** Orientation distribution of TT1-twinned grains, which were extracted from compressed Mg-7Y samples at all strain levels, overlaid with global Schmid factor (SF) calculation for different deformation systems. (a) SF for TT1 twinning system $\{10\bar{1}2\}\langle\bar{1}011\rangle$; (b) SF for pyramidal I <c+a> slip system; (c) SF for pyramidal II <c+a> slip system.

### 3.2.2 TT2-twinned grains

Extracted from the collected EBSD data for all compressed samples of the two alloys, 39 grains with TT2 twin formation were analyzed in total, which included 38 grains from Mg-7Y and 1 grain from Mg-1Y. Among this sample population, in 26 grains both types of tension twins were observed, which are thus referred to as TT1 and TT2 co-twinned grains. For the remaining 13 grains, only TT2 twins formed, which are thus referred to as TT2-only twinned grains. The crystallographic orientations of all TT2-twinned grains are shown in **Figure 10** in the form of an inverse pole figure with respect to the compression direction (*i.e.* extrusion direction). TT1 and TT2 co-twinned grains are colored in red, while TT2-only twinned grains are colored in black. **Figure 10** shows that the two groups of the TT2 twinned regions occupy different regions in the orientation space. The co-twinning region is concentrated near $[2\bar{1}10]$ pole, while the TT2-only grains occupied the more central region in the inverse pole figure with respect to the compression direction. **Figure 11** demonstrates the orientation distribution of all analyzed TT2-twinned grains overlaid with SF calculation for the two twinning systems. Based on the initial SF distribution, it is interesting to find that most of the TT1 and TT2 co-twinned grains (in red) could be favorable for the TT1 twinning system, while the majority of the TT2-only twinned grains (in black) could favor the TT2 twinning system.

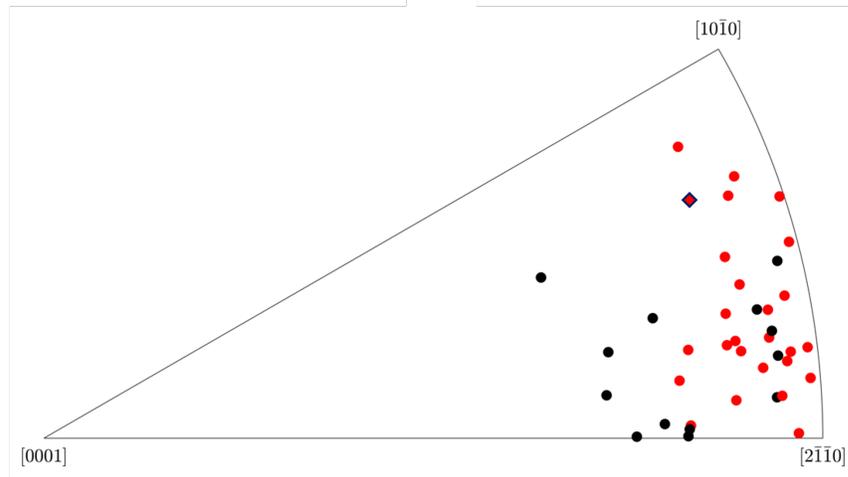

**Figure 10.** Orientation distribution of all TT2-twinned grains (39 grains in total were analyzed) in alloys Mg-1Y and Mg-7Y. One grain from Mg-1Y is marked in a diamond shape. 26 out of 39 grains are TT1 and TT2 co-twinned grains, which are marked in red. The remaining 13 grains are TT2-only twinned grains, which are marked in black. The inverse pole figure is generated from the compression direction (*i.e.* extrusion direction).

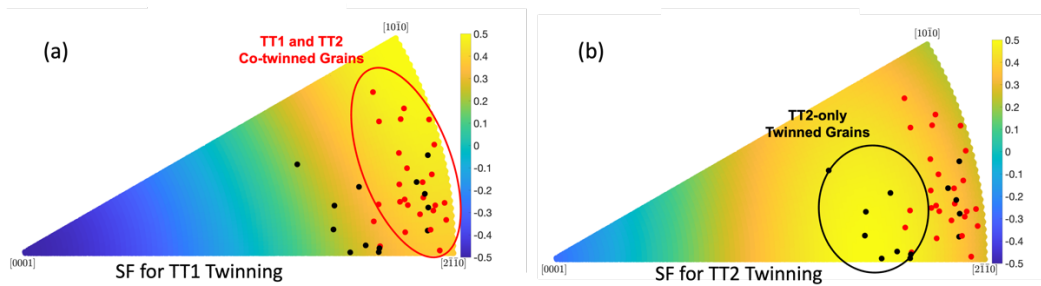

**Figure 11.** Orientation distribution of all analyzed TT2-twinned grains overlaid with global Schmid factor (SF) calculation for two twinning systems. (a) SF for TT1 twinning system; (b) SF for the TT2 twinning system.

The influence of compression strain level on the TT2 twin formation was also investigated. **Figure 12** exhibits the orientation distribution of all TT2-twinned grains in compressed Mg-7Y samples with different strain levels, which are overlaid with SF calculation for the two twinning systems. It is found that at the lower strain level of 5%, as shown in **Figure 12** (a) and (b), the majority of TT2-twinned grains were co-twinned by both TT1 and TT2 (colored in red), and their orientations could be favorable for TT1 twinning in terms of SF. In addition, with increasing the strain level to 8%, as shown in **Figure 12** (c) and

(d), more TT2-only twinned grains appeared (colored in black), and there were more grains with orientations favorable for TT2 twinning in terms of the calculated TT-2 SF.

The possible relation between the orientation of TT2-twinned grains and the activation of slip systems was explored as well. **Figure 13** shows the inverse pole figures, which contain the orientation distribution of all analyzed TT2-twinned grains, and the SF calculation results for different slip systems. It is interesting to notice that the orientation of TT2-twinned grains could also be favorable for pyramidal slip systems, especially grains preferentially oriented for Pyramidal Type II slip, in terms of the calculated SF, as shown in **Figure 13** (a) and (b).

The above experimental observations and analysis are summarized as follows. Both Mg-1Y and Mg-7Y alloys exhibited TT1 tension twins, whereas TT2 twins were predominantly observed in alloy Mg-7Y and were rare in alloy Mg-1Y. Increasing Y concentration suppressed TT1 twin formation while promoting TT2 twin development, though TT2 twins remained less frequent overall. Morphologically, TT2 twins appeared longer and thinner than TT1 twins and occurred either as TT2-only or in combination with TT1 twins. Crystallographic orientation analysis and SF calculations indicated that TT1 twinning was favored in grains oriented for TT1 or ⟨c+a⟩ pyramidal slip, with strain enhancing twinning in grains oriented for <c+a> slip. TT2-only twins were associated with parent grain orientations favorable for TT2 twinning systems. Further analyses analysis suggested that ⟨c+a⟩ pyramidal slip may contribute to TT2 activation.

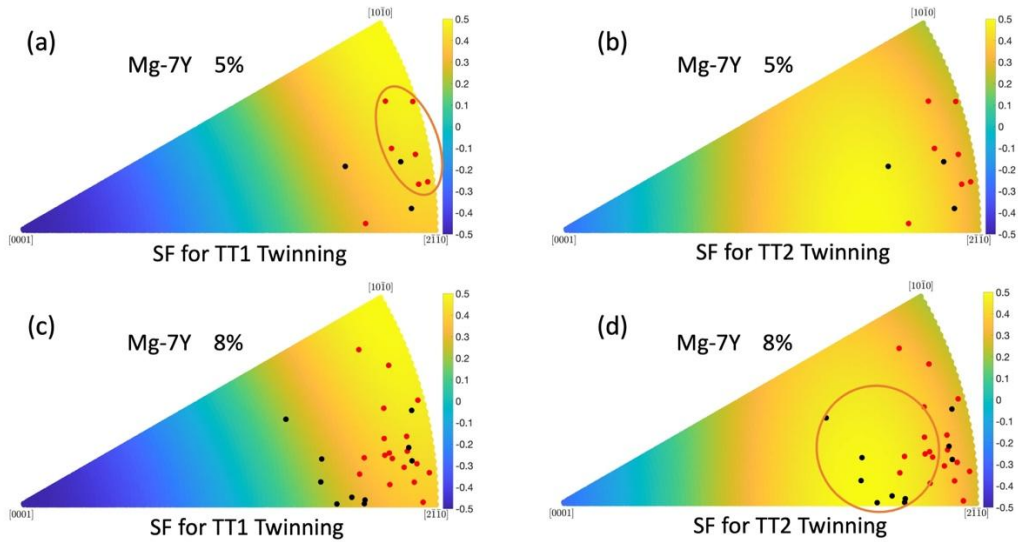

**Figure 12.** Orientation distribution of TT2-twinned grains in compressed Mg-7Y samples with different strain levels overlaid with global Schmid factor (SF) calculation for two twinning systems. (a) 5% strain compressed Mg-7Y, overlaid with SF for TT1 twinning system; (b) 5% strain compressed Mg-7Y, SF for TT2 twinning system; (c) 8% strain compressed Mg-7Y, overlaid with SF for TT1 twinning system; (d) 8% strain compressed Mg-7Y, SF for TT2 twinning system.

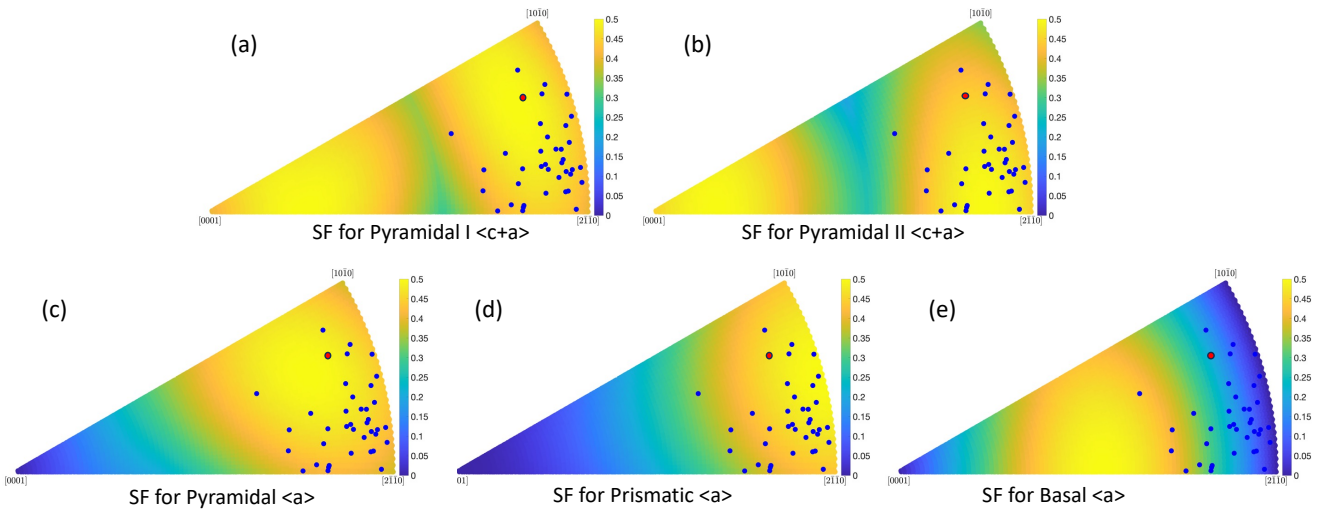

**Figure 13.** Orientation distribution of all analyzed TT2-twinned grains in Mg-1Y (one red point) and Mg-7Y (remaining dark blue points) overlaid with global Schmid factor (SF) calculation for different slip systems. (a) SF for pyramidal I <c+a> slip system; (b) SF for pyramidal II <c+a> slip system; (c) SF for pyramidal <a> slip system; (d) SF for prismatic <a> slip system; (e) SF for basal <a> slip system.

### 3.3 Crystal plasticity analysis of twin evolution

#### 3.3.1. CRSS calibration and its evolution with Y content

Using the procedure described in Section 2.2.3, constitutive parameters were assessed, and the resulting flow behaviors are compared with experimental measurements for Mg-1Y and Mg-5Y alloys are shown in **Figure 14** (a) and (b). Crystal plasticity parameters were also calibrated using the flow stress and twin evolution during compression deformation for Mg-5wt.%Y (1.42 at.% Y) and Mg-10wt.%Y (2.95 at.% Y) as reported in the literature [27]. The initial material processing for Mg-5Y and Mg-10Y was different from that for the conditions described in Section X for Mg-1Y and Mg-7Y. Hence, the constitutive parameters for Mg-5Y and Mg-10Y alloys may have additional variations that are not taken into account in this work. Nevertheless, both alloys provide relative trends in the CRSS and the field evolution with increasing Y content. The assessed CPFE model parameters for all four alloy systems are listed in Table II. Since the TT2 twinning is not active in the Mg-1Y and Mg-5Y alloys, the CRSS for twinning systems was established such that TT2 twinning is suppressed for those alloys.

Based on the observed experimental evidence, Stanford et al. [27] suggested a schematic crossover of CRSS of the TT1 and TT2 twins as a function of the Y concentration. The assessed initial CRSS ($s_0^\alpha$) values do follow that suggested trend, as shown in **Figure 15** (a), however, the variation of the assessed CRSS is non-linear. **Figure 15**(b) presents the assessed CRSS values for the active slip systems as a function of Y concentration, which also increase monotonically with Y addition. Thus, not only the magnitude of the CRSS but also the relative ratio of the CRSS for the different deformation modes also change with the Y addition.

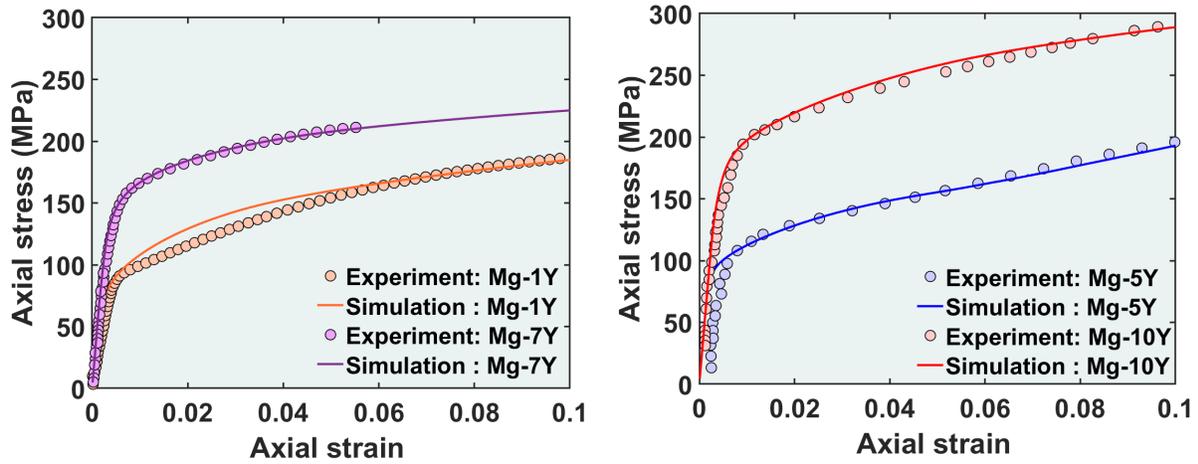
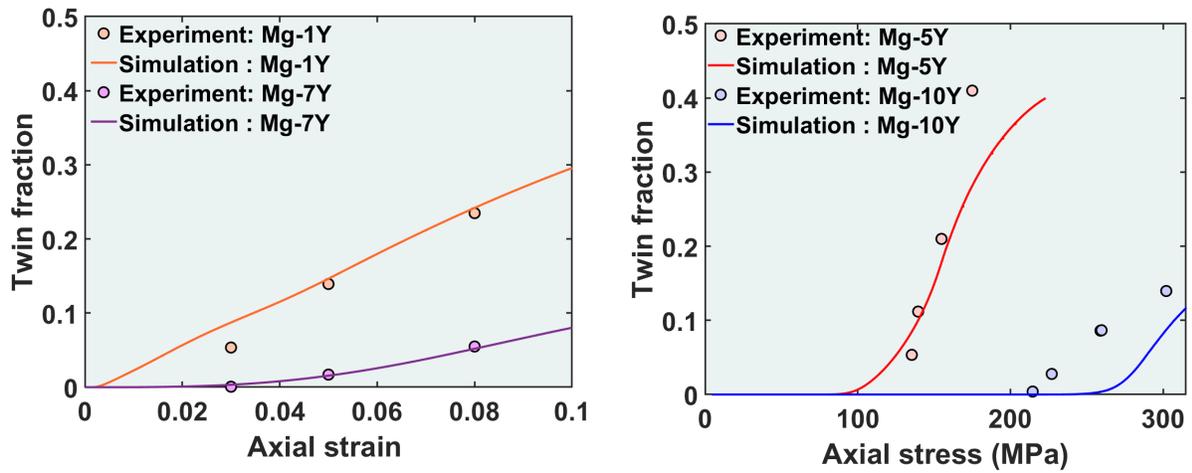

**Figure 14.** (a) Experimental flow behavior for Mg-Y alloys, along with the CPFE predicted flow curves. (b) Twin evolution for Mg-Y alloys and corresponding CPFE predictions.

Table II: Assessed CPFE Model parameters for the Mg-Y alloys

| | Material | $s_0^\alpha$ (MPa) | $h_0^\alpha$ (MPa) | $s_s^\alpha$ (Mpa) | $\alpha^\beta$ |
|---|---|---|---|---|---|
| Basal | Mg1Y | 10 | 1000 | 35 | 1 |
| | Mg5Y | 15 | 1000 | 40 | 1 |
| | Mg7Y | 22 | 1000 | 40 | 1 |
| | Mg10Y | 30 | 1000 | 60 | 1 |
| Prismatic | Mg1Y | 36 | 1000 | 200 | 6 |
| | Mg5Y | 40 | 1000 | 200 | 6 |
| | Mg7Y | 70 | 1000 | 200 | 6 |
| | Mg10Y | 90 | 1000 | 200 | 6 |
| Pyramidal <a> | Mg1Y | 60 | 1000 | 500 | 6 |
| | Mg5Y | 70 | 1000 | 500 | 6 |
| | Mg7Y | 92 | 1000 | 500 | 6 |
| | Mg10Y | 110 | 1000 | 600 | 6 |
| Pyramidal <c+a> | Mg1Y | 60 | 1000 | 500 | 6 |
| | Mg5Y | 75 | 1000 | 500 | 6 |
| | Mg7Y | 102 | 1000 | 500 | 6 |
| | Mg10Y | 120 | 1000 | 600 | 6 |
| Tensile Twin I | Mg1Y | 45 | 500 | 70 | 1 |
| | Mg5Y | 60 | 500 | 85 | 1 |
| | Mg7Y | 135 | 500 | 143 | 1 |
| | Mg10Y | 175 | 500 | 180 | 1 |
| Tensile Twin II | Mg1Y | 80 | 500 | 250 | 1 |
| | Mg5Y | 80 | 500 | 250 | 1 |
| | Mg7Y | 82 | 500 | 102 | 1 |
| | Mg10Y | 100 | 500 | 120 | 1 |

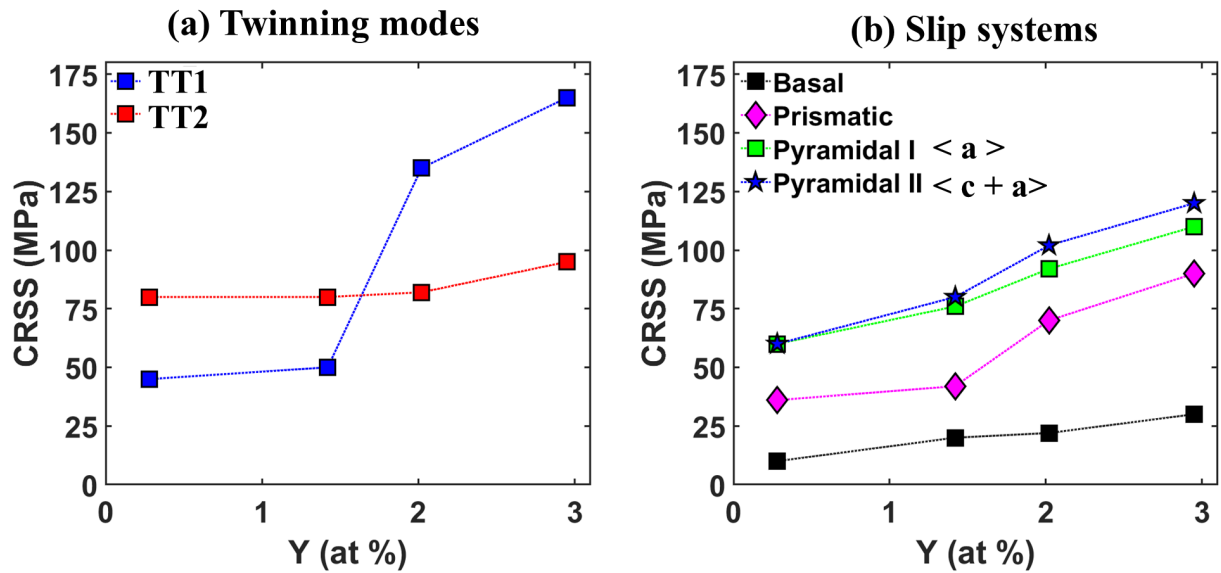

**Figure 15.** (a) Assessed initial twin CRSS as a function of the Y concentration for the TT1 and TT2 twin systems, showing a change in relative strength of the two twin systems with increasing Y concentration. (b) Assessed CRSS for the basal, prismatic, pyramidal <a> and pyramidal II <c+a> slip systems with increasing the Y concentration.

Many HEDM and micro-pillar compression experiments in the literature have characterized CRSS for the active deformation modes in Mg-Y alloys. Based on the HEDM experiments, Huang et al. observed that the ratio of the prismatic to basal CRSS varied from 1.8 to 2.7, and pyramidal <a> to basal CRSS varied from 1.6 to 1.8 for Mg-5wt.%Y alloy [50]. Similarly, Wang et al. characterized basal, prismatic, and pyramidal <a> CRSS equal to 12 MPa, 38 MPa, and 36 MPa, respectively, for Mg-3wt.%Y [48]. Both basal and prismatic CRSS from the HEDM observations fall within the range of the assessed values determined here for Mg-1Y and Mg-5Y, while the assessed pyramidal <a> CRSS for all of the alloys in the current study are higher than the observations reported [48,50].

In addition, Wu et al. analyzed the deformation mechanisms and corresponding CRSS values using micropillar compression tests of single crystals for Mg-0.4wt.%Y and Mg-4wt.% Y [51]. The authors measured the twin CRSS for Mg-0.4Y≈ 45 MPa and for Mg-4Y≈113 MPa. Li et al observed twin CRSS of ~38 MPa for Mg-2wt.%Y [52], while Yu et al observed twin CRSS of ≈148 MPa for Mg-5wt.%Y-0.08wt.%Ca [53], again pointing out the twin CRSS increment with Y addition. Similarly, an eight-fold increment in CRSS was reported for the Mg-Gd alloys [54]. Since micropillar compression tests eliminate grain boundary nucleation effects, some overestimation in CRSS values is expected from the micropillar tests. Keeping that in mind, the assessed CRSS variations we report in **Figure 15** are reasonably consistent with the observed trends in these previous studies.

For most of the twinning modes in HCP metals, atomic shear alone does not bring the atoms to the equilibrium locations of the twin, and hence, atomic shuffle is also required. For the TT1 twin, atomic shuffle movements along the twin directions, as well as perpendicular to the twin directions, are required. Therefore, the addition of the solute atoms with a higher atomic radius, like Y, can highly influence the atomic shuffle movements, resulting in a higher twin CRSS with higher solute content [27]. On the other hand, as mentioned in the introduction, the TT2 twin does not require atomic shuffle movement along the shear direction; hence, the influence of Y on TT2-CRSS can be lower. As described earlier, recent atomistic studies suggest that a difficult out-of-plane shuffle perpendicular to the twin direction may, however, be necessary for TT2 twin formation, which can influence its critical twin growth [29]. This can be correlated with higher CRSS for the TT2 compared to the TT1 twin in Mg alloys with a lower Y concentration. As shown in Figure 16a, the change of twin CRSS with Y addition is not constant, for the TT1 twin, there appears to be an abrupt change in the assessed CRSS between 1.5at% and 2 at% Y, while for the TT2, there is a gradual increase in the assessed CRSS with increasing Y content in the solid solution.

In addition, the twin boundary structures of the two Mg-Y alloys studied here were characterized using high-resolution scanning transmission electron microscopy (STEM). Figure 16 presents STEM-HAADF images of the regions near TT1 twin boundaries after compression deformation. Notably, significant Y segregation was detected at the twin boundaries in Mg-7Y (Figure 16 (b)), where bright-contrast clusters of Y were distributed along the boundary. In contrast, no such segregation was observed in Mg-1Y (Figure 16 (a)). Both coherent twin boundary (CTB) and basal-prismatic (BP) interface structures were identified, with Mg-7Y showing a higher prevalence of BP steps, which we suggest are preferential locations for Y to segregate (Figure 16(c)). Similar deformation-induced segregation at room temperature, without annealing treatment, has previously been reported in Mg-Y alloys with higher solute concentrations [55]. It is likely that the segregation of Y at twin boundaries contributes to the increase in CRSS, particularly the sharp rise in CRSS for TT1 twins when Y content approaches ~2 at.% (Figure 15 (a)).

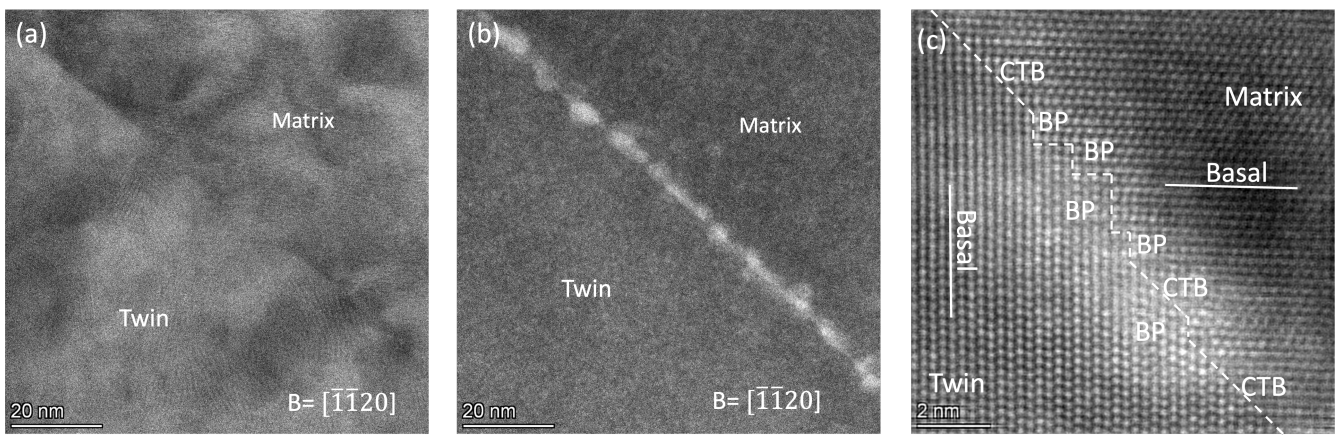

**Figure 16.** STEM-HAADF images taken with $[\bar{1}\bar{1}20]$-Mg zone axis from Mg-1Y (a) and Mg-7Y alloy (b and c) after compression deformation with ~5% plastic strain. (a) and (b) Low magnification; (c) High magnification to show the twin boundary structure and Y distribution in the TT1 twin boundary in Mg-7Y.

In the present study, we analyzed CRSS and twin activity changes as functions of Y content in the solid solution across varying alloy compositions. The same influence can be studied by comparing alloys in different heat-treated conditions, as the solute content in the matrix changes. For example, Li et al. [28] analyzed twin activity in the magnesium alloy GW103 (Mg-10Gd-3Y-0.5Zr wt%) in two different heat-treated conditions, T4 and T6 (peak-aged). RE solute content in the matrix is expected to be higher for T4 heat treated conditions compared to T6 heat treated conditions. Li et al. [28] showed that the TT2 twin evolved before the TT1 twins in the T4 heat-treated GW103 [28], denoting early critical nucleation and growth of TT2 twins compared to TT1 twins. On the other hand, the TT1 twin evolved before the TT2 twin in the T6 condition, which would be expected to have a lower RE solute content in the solid solution [28]. This supports the current study's results, which indicate that the presence of higher RE elements in the solid solution facilitates TT2 twin formation. Similarly, Lentz et al also observed a higher twin fraction in an age-hardened Mg-Y-Nd (WE54) alloy compared to as-extruded alloys [56]. The authors observed precipitate formation, resulting in a decrease in solute content and a corresponding increase in twin activity in the age-hardened alloy matrix. In this regard, Lentz et al. [56] also corroborated a decrease in the TT1 CRSS from 115 MPa at the as-extruded state (higher solute content relative to age-hardened condition) to 78 MPa in the age-hardened state, which is also in accordance with **Figure 15**.

Although the crystal plasticity results alone cannot support or negate any specific atomistic influence, our results do show a higher assessed CRSS increment for the TT1 twinning with the Y addition. We also note that with inverse multi-parameter optimization, the assessed CRSS values are not necessarily unique and must be verified by additional studies. Nonetheless, these results provide important insights into the changes in the relative strength of different deformation modes with Y additions. Next, we will consider the relative strength variations of the twinning and slip systems in terms of the plastic anisotropic measure.

Recently, the plastic anisotropic measure (PA) was introduced by Kumar et al. [57] as,

$$PA = \frac{\tau_0^{Prismatic} - \tau_0^{Basal}}{\tau_0^{Twin} - \tau_0^{Basal}} \qquad (12)$$

Based on the definition, a low PA measure indicates a higher propensity for slip activation, while a high PA measure indicates a higher propensity for twin deformation. Kumar et al. [57] categorized alloys into two groups: alloys with PA< 2 and alloys with PA>2. The alloys in the first category, therefore, exhibit a lower twinning propensity. **Figure 17** (a) shows the PA measure for the Mg-Y alloys considering both the twin types. The arrows in the subfigure indicate the direction of change in the PA measure as the Y content increases. It is apparent that the Y addition results in a decrease in the PA measure with respect to the TT1 twin, correlating well with a lower twin fraction and an increase in Y concentration, as shown in **Figure 4** and **Figure 5**. Kumar et al [58] also associated lower PA measure with difficult twin thickening and twin transmission using an explicit twin propagation model. Thus, a lower PA measure indicates difficult twin growth as well as twin propagation of TT1 twins with increasing %Y in Mg alloys. Interestingly, assessed CRSS values show an opposite variation of PA measure for the TT2 twinning, indicating an increasing propensity for the TT2 twin formation with increasing Y content.

The PA measure only considers the relative resistance of the prismatic and twin systems compared to basal slip systems. However, many experimental observations have noted increased pyramidal slip activity with the addition of Y [12]. Therefore, **Figure 17** (b) shows the relative ratio of the pyramidal II CRSS to the twin CRSS. Higher pyramidal/twin CRSS ratios indicate that twin resistance is lower compared to the pyramidal slip, favoring increased twin activation. Similar to the PA measure, it can be noted that the relative resistance to activation for pyramidal slip systems with respect to the TT1 twin decreases with increasing Y concentration. On the other hand, the CRSS ratio increases for the TT2 twin. Similar observation of the decrease in non-basal slip to twin CRSS ratio with increasing the Nd concentration was reported by [59]. The trends in both the PA measure and the pyramidal to twin CRSS ratio indicate a decrease in the TT1 twin and an increase in TT2 twin activity with increasing %Y. It is therefore apparent that with increasing Y content, the TT1 twin activity decreases with the corresponding increase in the prismatic, pyramidal slip activation.

We also note that both twin nucleation and growth are affected by higher RE solutes. He et al. [60] demonstrated that a larger atomic radius, such as that of Y in Mg, reduces the TT2 twin boundary energy by relaxing elastic strains along the boundary. This influences both the ease of twin formation and the subsequent growth of TT2 twins in Mg-RE alloys. While the CPFE-assessed CRSS does not explicitly separate the effects of twin nucleation and growth kinetics, the relative trends in CRSS reflect the overall influence of solute Y on twin evolution in Mg-Y alloys.

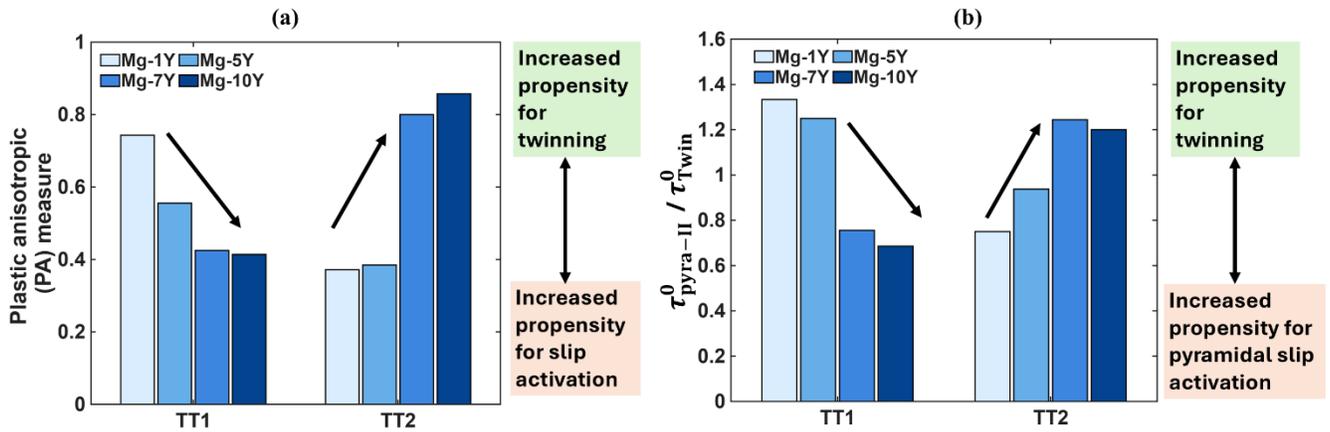

**Figure 17.** (a) Plastic anisotropic (PA) measure defined in Eq. (12) and (b) ratio of the CRSS for pyramidal II slip system to twinning system for the TT1 and TT2 twinning systems for all the Mg-Y alloys considered in this work. As the CRSS varies for all the considered deformation modes, the relative ratio of the CRSS for deformation modes also changes. Arrows in both subfigures indicate the general direction of changes in the deformation mode with increasing Y content. Both subfigures illustrate that, based on the relative values of CRSS, with the addition of yttrium in Mg alloys, the TT1 twin activation becomes more difficult while the activation of the TT2 twin becomes easier.

### 3.3.2 Twin activity and local stress-strain evolution

An increase in the Y concentration leads to an increase in the relative volume fraction of the TT2 twins. As the twin shear for the TT2 twin is nearly five times that of the TT1 twin, even with a smaller volume fraction, TT2 twin shear accommodation can also become important. **Figure 18** presents the amount of total shear strain accommodation of both twins in Mg-Y alloys, determined using $\gamma^{twin} = f^{twin} S^{twin}$ ($S^{twin}$ represents characterstic twin shear). In Mg-7Y alloy, since the TT2 twin fraction is significantly lower than that of the TT1, shear strain is mainly accommodated by TT1 twins. On the other hand, similar shear strain accommodation by both twin types can be observed in Mg-10Y at larger strain levels. The total shear strain accommodated by the TT2 twin locations is even higher than that of the TT1 twin locations at 10% strain. Greater shear strain accommodation by the TT2 twin can influence the evolution of stress and strain at these locations. Hence, we will focus on the local field evolution associated with the TT1 and TT2 twin activities.

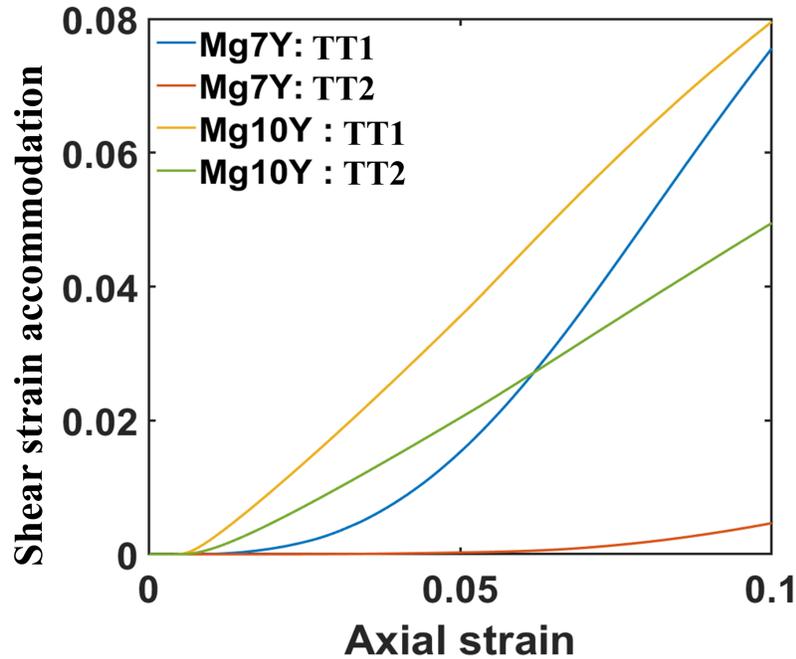

**Figure 18.** Total shear strain accommodated by the TT1 and TT2 twinning during deformation.

To understand the spatial correlations between strain accumulation and twin evolution, we first consider an idealized tri-crystalline microstructure in CPFE analysis, as shown in **Figure 19** (a). Consideration of such a tricrystalline geometry facilitates understanding of grain neighborhoods and orientation relations in a simplified microstructure. Thus, the influence of the complex boundary conditions and grain shape is not included in such an analysis. The central grain was assigned two different orientations, which showed the TT1 and TT2 twinning in the Mg-7Y alloy from **Figure 6**. **Figure 19** (b) shows two randomly selected grains, designated as G1 and G2, along with their grain neighborhoods.

**Figure 19** (c) shows the orientations of the selected grains and their neighboring orientations in the (0001) pole figure. To further understand the orientations of the selected neighborhood, **Figure 19** (d-e) also shows Schmid factors of the grains for basal slip, TT1 and TT2 twins. As expected, the G1 grain has the highest Schmid factor for the TT1 twin while the G2 grain shows the highest Schmid factor for the TT2 twin. The G1 grain is surrounded by orientations with high Schmid factors for basal slip and low Schmid factors for TT1 twins. On the other hand, the G2 grain is surrounded by orientations with low Schmid factors for basal slip and high Schmid factors for TT1 twins. These cases represent realistic, simplified depictions of the influence of grain neighborhoods during deformation.

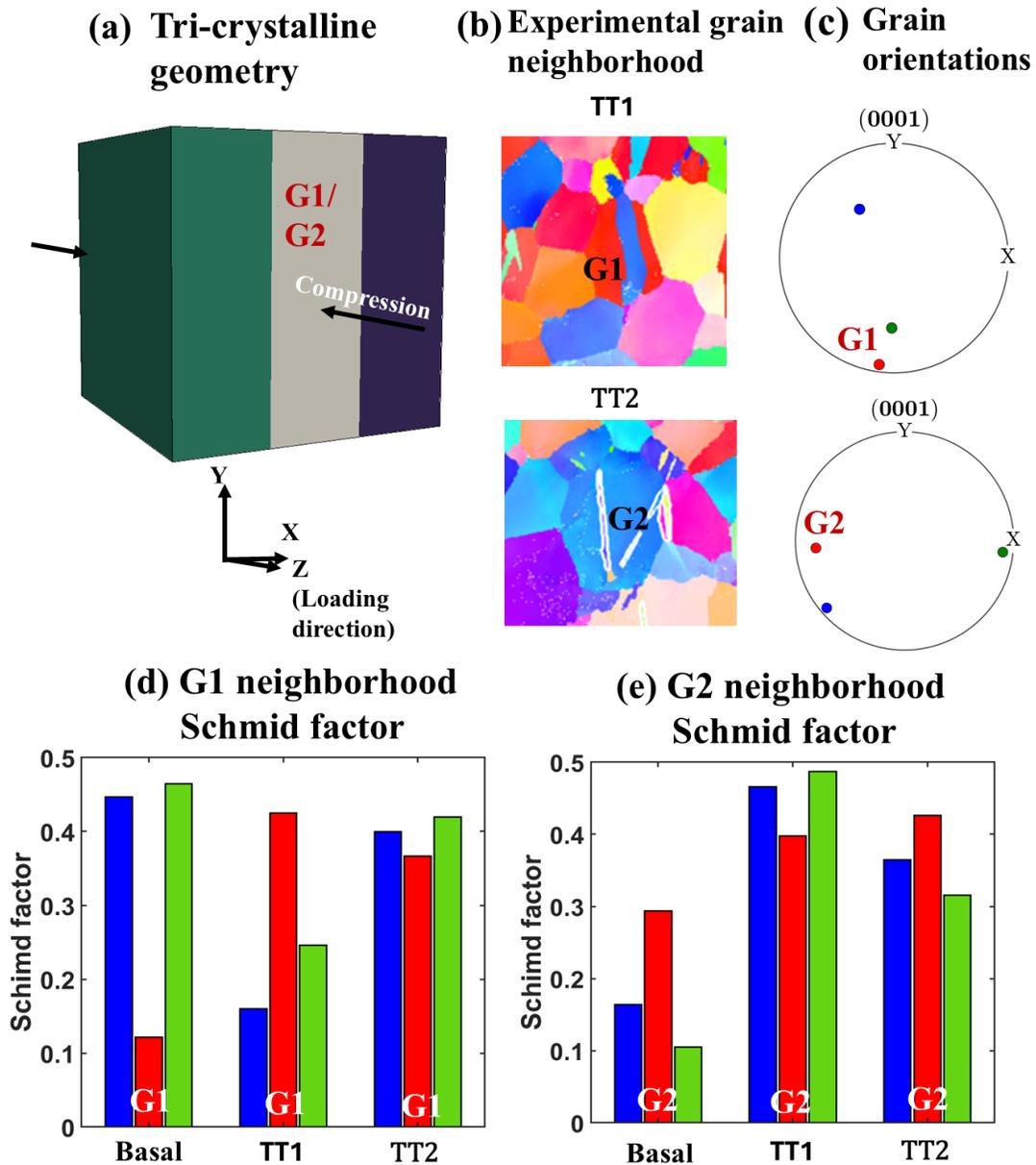

**Figure 19.** (a) Initial tri-crystalline microstructure utilized for the deformation simulation. The central grain is denoted by G1/G2 and is assigned the parent grain orientation from Figure 6. (b) Experimental parent grain orientations used for the central grain. The neighboring grain orientations were chosen randomly from the observed grain neighborhood. (c) The {0001} pole figure showing grain orientations assigned to trycrystalline geometry along with neighbor grain orientations. Schmid factors for basal slip, TT1 twin and TT2 twins are shown for the G1 (d) and G2 (e) parent grain orientation.

**Figure 20** presents equivalent stress and strain distributions at locations with more than 10% TT1 and TT2 twin activity in the central grains (G1 and G2) of interest shown in **Figure 19** (a). In Figure 20, the transparent tri-crystalline microstructure is overlaid on the twinned regions to offer a spatial view of the twin locations within the central grain. Even in this simple geometry, the figure consistently shows a lower volume fraction of TT2 twin activity (in G2) compared to TT1 twin activity (in G1). We observe heterogeneous stress and strain distributions for the twin locations. The locally constrained active TT2 twin region (in G2) does show a relatively higher strain accumulation compared to the TT1 twin (in G1). The opposite trend was observed for the stress distributions. Some of the high-stress and strain locations are highlighted with black arrows for the illustration. The ratio of the mean von Mises strains at the TT1 twin locations to the mean von Mises strain of the grain G1 is 1.37, while the same ratio for the TT2 twin locations in the grain G2 is 1.74. Despite differences in the neighborhoods, Schmid factors, and spatial locations, higher strain accommodation in the constrained volume for the TT2 twin is observed in the simple tri-crystalline geometry.

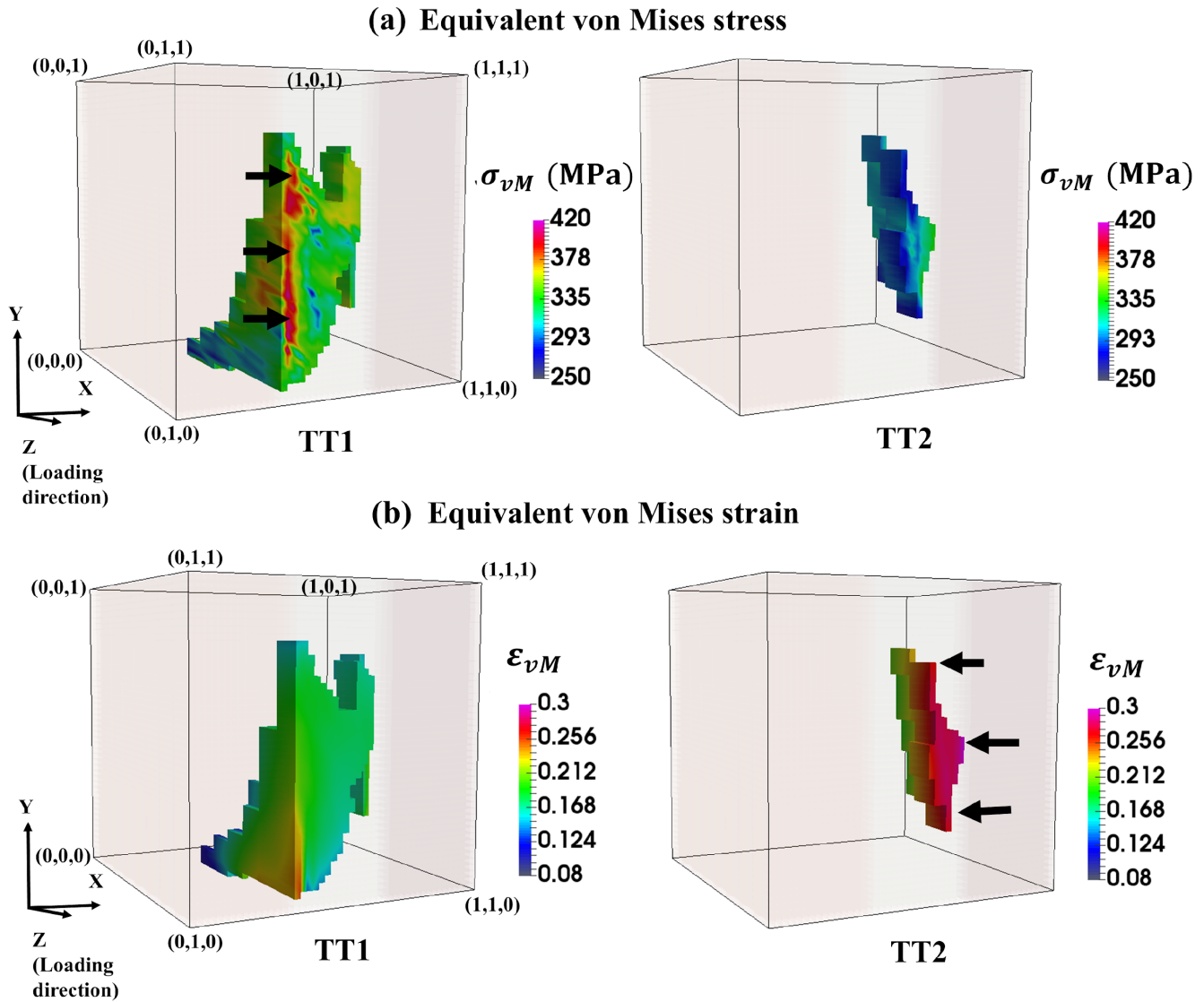

**Figure 20.** Calculated equivalent von Mises (a) stress and (b) strain distribution at the locations showing the TT1 and the TT2 twin activity in the G1 and G2 grains from Fig. 20a. Black arrows in each subfigure highlight high stress and strain locations. Higher stress accumulation at the TT1 twin locations and higher strain accumulation at the TT2 twin locations can be observed.

We also analyzed the strain and stress distribution for the entire RVE. **Figure 21** presents the distribution of the axial stress and strain for locations with the TT1 and TT2 twin activity. As the nominal values are different for the two alloys, both stress and strain distributions are normalized by their respective mean values within the RVE. The mean values of the distributions are also marked in the subfigures. Such statistical distributions provide many insights into the local stress and strain evolution of the two different twins. First, the stress distributions for both twins showed smaller deviations from the mean values, while the strain distributions showed larger extreme strain deviations in both alloys. In the case of von Mises stresses, the histogram for the TT2 twin is shifted towards lower stress values compared to the TT1 twin locations. The mean value of the distribution for the TT2 twin locations is slightly lower than one, while the mean value for the TT1 twin locations is more than one.

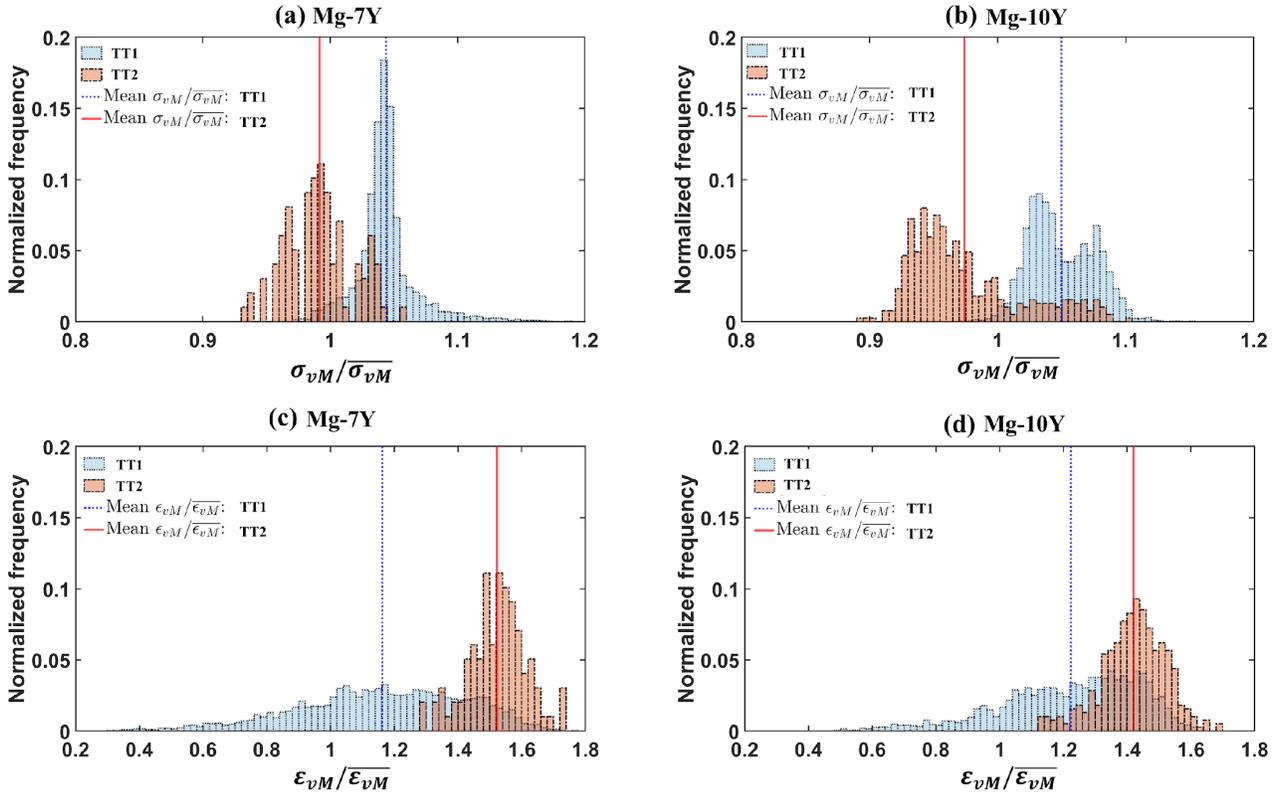

**Figure 21.** Histogram of von Mises stresses ($\sigma_{vM}$) at twin locations normalized by the mean stress of the RVE ($\bar{\sigma}_{vM}$) for (a) Mg-7Y (b) Mg-10Y. Histogram of von Mises strains ($\varepsilon_{vM}$) at twin locations normalized by the mean stress of the RVE ($\bar{\varepsilon}_{vM}$) ) for (a) Mg-7Y (b) Mg-10Y. The mean values for both the twin types are also marked by the blue and red lines in each subfigure. The strain distribution is wider than the stress distribution for both alloys. A clear shift of the stress histogram for the TT1 twin and the strain histogram for the TT2 twin can be observed.

We also note that the mean for strain distributions is higher than the average strain of the entire RVE for both the twin types. The TT1 twin locations show a much wider strain distribution compared to the TT2 twin locations. Most of the strain distribution of the TT2 twin locations is located towards the right of unity, indicating higher local strain accommodation than the average strain of the RVE. Thus, both the simple tri-crystalline geometry and the polycrystalline RVE illustrate that the local strain is higher for the TT2 locations relative to the TT1 twin locations. Such higher strain evolution can influence the damage accumulation during the deformation. High strain accumulated at individual twin boundaries can act as a nucleation site for crack formation, twin nucleation in adjacent grains or parent grains, or even recrystallization [25,61]. For example, Chen et al observed the presence of the TT1 twins on the boundary of the TT2 twin, indicating that the TT2 twin boundary can act as a nucleation site for further twin variants in WE43 alloy [23]. Wang et al. observed visible damage near the TT2 twin during the bend test of the Ti alloy [62].

To understand the effect of different stress states on the damage initiation, we also studied the stress triaxiality of the RVE. During heterogeneous deformation, the stress state can be characterized by the stress triaxiality ratio. High stress triaxiality indicates a propensity for void nucleation, growth, and coalescence, which initiates damage [63,64]. Stress triaxiality is defined as the ratio of the hydrostatic stress ($\sigma_h$) to von Mises equivalent stress as,

$$stress\ triaxality = \frac{\sigma_h}{\sigma_{vM}} \qquad (13)$$

$$\sigma_h = tr(\sigma) \qquad (14)$$

**Figure 22** presents the cumulative distribution of stress triaxiality for Mg-7Y and Mg-10Y. Both twin types exhibit extreme values, with stress triaxiality exceeding 1, indicating a high potential for damage initiation. Comparatively, the TT1 twin shows a broader distribution with larger extreme values compared to the TT2 twin. The stress state was not significantly different for either twin type based on the distribution of stress triaxiality. However, once initiated, higher strain accumulation at the TT2 twin locations can influence the further propagation of damage and material failure.

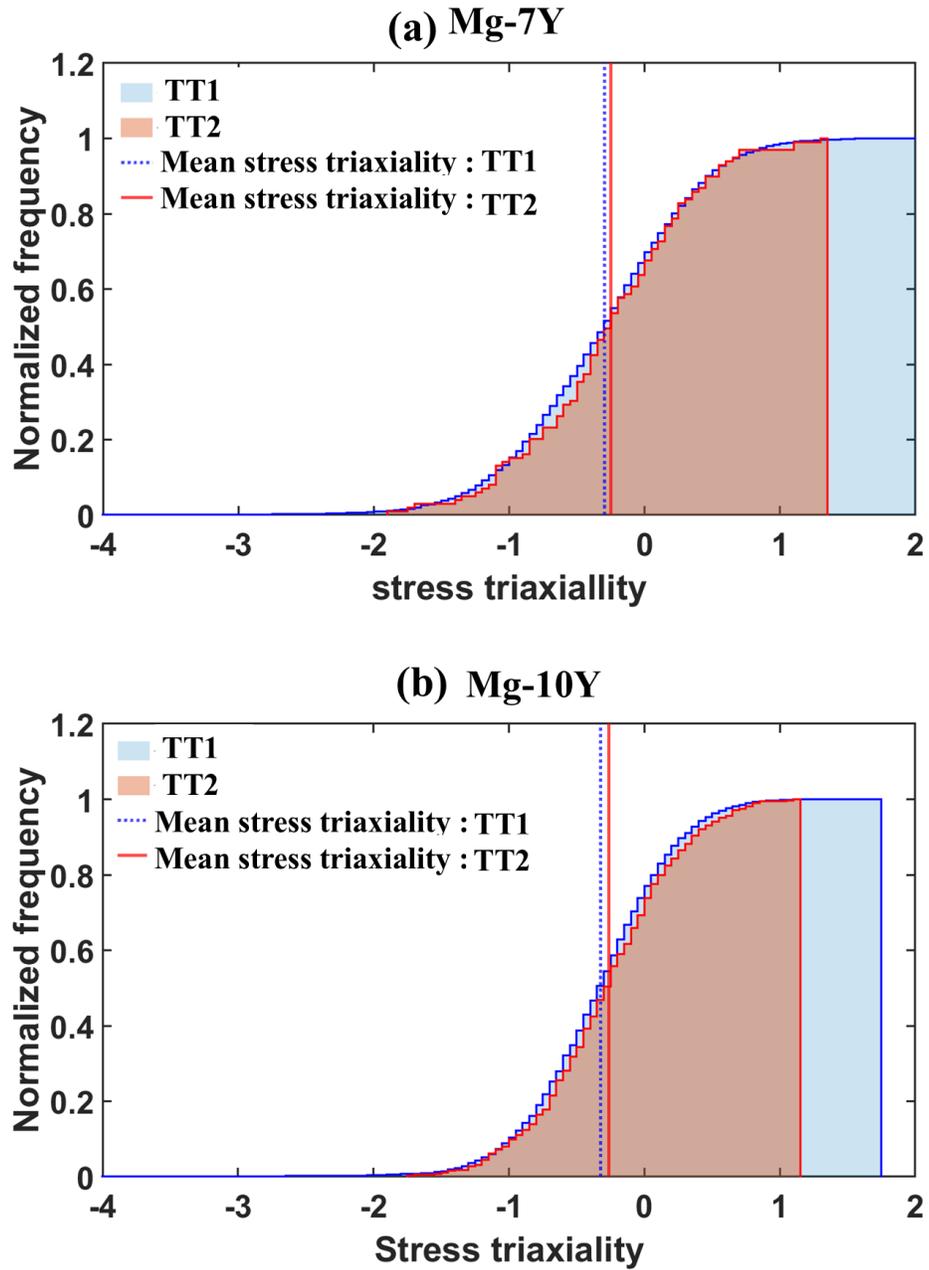

**Figure 22.** Stress triaxiality calculated at 10% axial strain during compression deformation for (a) Mg-7Y and (b) Mg-10Y.

We note that the current study does not consider the influence of twin interaction with grain boundaries or twin-twin interactions and double twinning, which can also influence the stress state and local mechanical evolution during the deformation. Although this work focuses on the twin activity in Mg-Y alloys, the observed influence of twin activation, CRSS evolution, and field distributions can also be applied to understand the impact of other alloy additions. It is crucial to understand the activation of non-basal slip, TT1, and TT2 tension twins, and to design Mg-RE alloys to achieve an optimal balance between high strength and reasonable ductility. Hence, the statistical overview of tension twin activation presented in this work can be useful for new alloy development and for understanding twin evolution under different deformation conditions in Mg alloys.

## 4 Summary and Conclusions

This study investigated the evolution of tension twins during the compression of different Mg-Y alloys using both experimental and finite element-based crystal plasticity analyses. The simulations considered potential deformation modes, elastic anisotropy, and two types of tension twins, ensuring that all key changes in material properties resulting from Y additions are considered. Previous research has primarily focused on low-Y% Mg alloys or on single-crystal deformation. In contrast, this study presents a comprehensive and systematic analysis of TT1 and TT2 twinning in Mg alloys with higher rare-earth solutes, such as 7-10 at% Y. The extensive experimental analysis offers novel insights into the correlation between TT1 and pyramidal slip at higher Y content, as indicated by Schmid factor analysis. Changes in the twinning propensity with Y addition were rationalized in terms of relative changes in the critical resolved shear stresses of the active slip and twin modes. Moreover, we demonstrate that TT2 twins can accommodate larger shear strains in smaller volume fractions, resulting in localized high-strain areas that may influence damage or promote further secondary twinning. These findings could benefit both the design and application of advanced Mg alloys. The main conclusions are as follows.

1. TT1-$\{10\bar{1}2\}\langle\bar{1}011\rangle$ tension twins formed in both alloys, with low and high Y concentrations. TT2-

$\{11\bar{2}1\}\langle\bar{1}\bar{1}26\rangle$ tension twins were observed in alloy Mg-7Y but were rarely observed in alloy Mg-1Y. Higher Y concentrations decreased the formation of TT1 twins and promoted the formation of TT2 twins.

2. Relative to TT1 twins, the fraction of TT2 twins formed in Mg-7Y was significantly lower. TT2 twins appeared longer and thinner than TT1 twins. TT2 twins formed explicitly in some grains, whereas both TT1 and TT2 twins were observed in many co-twinned grains.

3. TT1 twinned grains were categorized into two grain groups according to the grain crystallographic orientation analysis. The first group (Group A) is the predominant group and includes TT1-twinned grains that are favorable for the TT1 twinning system from the perspective of the global Schmid factor. Meantime, the second group (Group B) are grains in which TT1 twins were observed and which were also favorable for the pyramidal <c+a> slip system. The formation of the TT1 twins in group B grains was promoted with increasing compression strain.

4. The majority of TT2-only twinned grains were favorable for the TT2 twinning system in terms of global Schmid factor, while most of the TT1 and TT2 co-twinned grains were favorable for the TT1 twinning system. Higher compression strain promoted the formation of TT2-only twins.

5. Using a rigorous and systematic CPFE calibration process, assessments were made of the CRSS and hardening parameters for four Mg-Y alloys. These results show a monotonic increase in both the slip and twin resistance with Y concentration. The relative ratio of prismatic/pyramidal slip to TT1 twin CRSS decreases with Y concentration, resulting in easier slip and more difficult TT1 twin activation. In contrast, the relative ratio of the prismatic/pyramidal slip to TT2 twin CRSS increases with Y concentration demonstrating a higher propensity of TT2 twinning. We note that the volume fraction of TT1 twins is still higher than that of the TT2 twin up to 10 wt% Y in Mg.

6. As the characteristic twin shear is nearly five times higher than that of the TT2 twin, it can

accommodate larger shear strain in smaller volume fractions. In both a simple tri-crystalline geometry and the polycrystalline RVE, the TT2 twin locations exhibit high localized strain concentrations.

7. The TT1 twin locations show a wider strain distribution compared to the TT2 twin locations. The mean for strain distribution for both the twin types is higher than the average of the whole RVE, signifying higher strain accumulation at the twin locations. The cumulative stress triaxiality distribution for both the twin types is similar, with a broader distribution for the TT1 twin locations. Higher strain accumulation at both the twin locations can further influence damage propagation, recrystallization, or twin nucleation.

**Data availability**

The experimental and simulation data supporting this study are available on Materials Commons at xxx. [DOI to be established upon article acceptance]

**Disclosure statement**

The authors declare that they have no known competing financial interests or personal relationships that could have appeared to influence the work reported in this paper.

**Acknowledgement**

This work was supported by the PRISMS (PRedictive Integrated Structural Materials Science) center which is located at University of Michigan and funded by the U.S. Department of Energy, Office of Basic Energy Science, Division of Materials Science and Engineering (Grant award number DE-SC0008637). We employed computational resources provided by Advanced Research Computing at the University of Michigan, Ann Arbor, and Bridges-2 at Pittsburgh Supercomputing Center (allocation MSS160003 from the Advanced Cyberinfrastructure Coordination Ecosystem: Services & Support (ACCESS) program, which is supported by National Science Foundation grants #2138259, #2138286, #2138307, #2137603, and #2138296).